\documentclass[twocolumn,showpacs,superscriptaddress,aps,pre,10pt]{revtex4-1} 
\usepackage{epsfig,amsmath,amsfonts,bbold}

\usepackage{ulem}

\newcommand{\Laplace}{\mathop{\vcenter{\hbox{\Large$\mathcal{L}$}}}\nolimits}

\newcommand{\beq}{\begin{equation}}
\newcommand{\eeq}{\end{equation}}
\newcommand{\beqarr}{\begin{eqnarray}}
\newcommand{\eeqarr}{\end{eqnarray}}
\def\besplit#1\eesplit{\begin{equation}\begin{split}#1\end{split}\end{equation}}
\def\bsubeq#1\esubeq{\begin{subequations}\begin{align}#1\end{align}\end{subequations}}
\def\bal#1\eal{\begin{align}#1\end{align}}
\def\bmul#1\emul{\begin{multline}#1\end{multline}}
\newcommand\picpath{./}

\usepackage{color}

\begin{document}
\normalem

\title{Synchronization of weakly perturbed Markov chain oscillators}
\author{Ralf T\"onjes}
\affiliation{Ochadai Academic Production, Ochanomizu University, Tokyo 112-8610, Japan}
\author{Hiroshi Kori}
\affiliation{Ochadai Academic Production, Ochanomizu University, Tokyo 112-8610, Japan}
\affiliation{PRESTO, Japan Science and Technology Agency, Kawaguchi, Saitama 332-0012, Japan}

\begin{abstract}
Rate processes are simple and analytically tractable models for many dynamical systems that switch stochastically between a discrete set of quasi stationary states; however, they may also approximate continuous processes by coarse-grained, symbolic dynamics. In contrast to limit-cycle oscillators that are weakly perturbed by noise, in such systems, stochasticity may be strong, and topologies more complicated than a circle can be considered. Here, we apply a second-order time-dependent perturbation theory to derive expressions for the mean frequency and phase diffusion constant of discrete-state oscillators coupled or driven through weakly time-dependent transition rates. We also describe a method of global control to optimize the response of the mean frequency in complex transition networks.
\end{abstract}
\pacs{05.45.Xt, 02.50.Ga, 82.40.Bj, 05.40.-a}
\maketitle
\section{Introduction}
\begin{figure}[!t] 
%
 \includegraphics[width=8cm]{\picpath/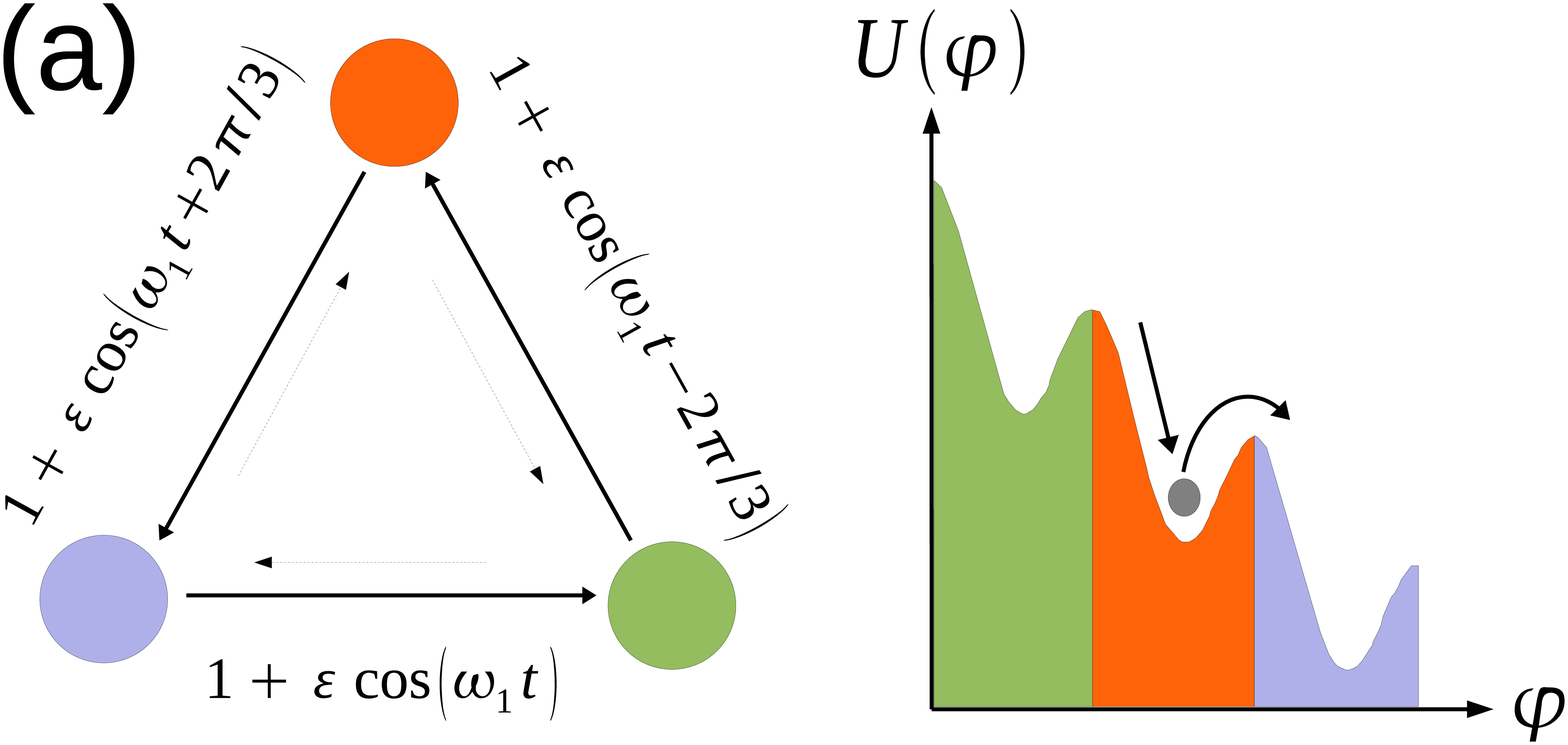} 
 \includegraphics[width=8cm]{\picpath/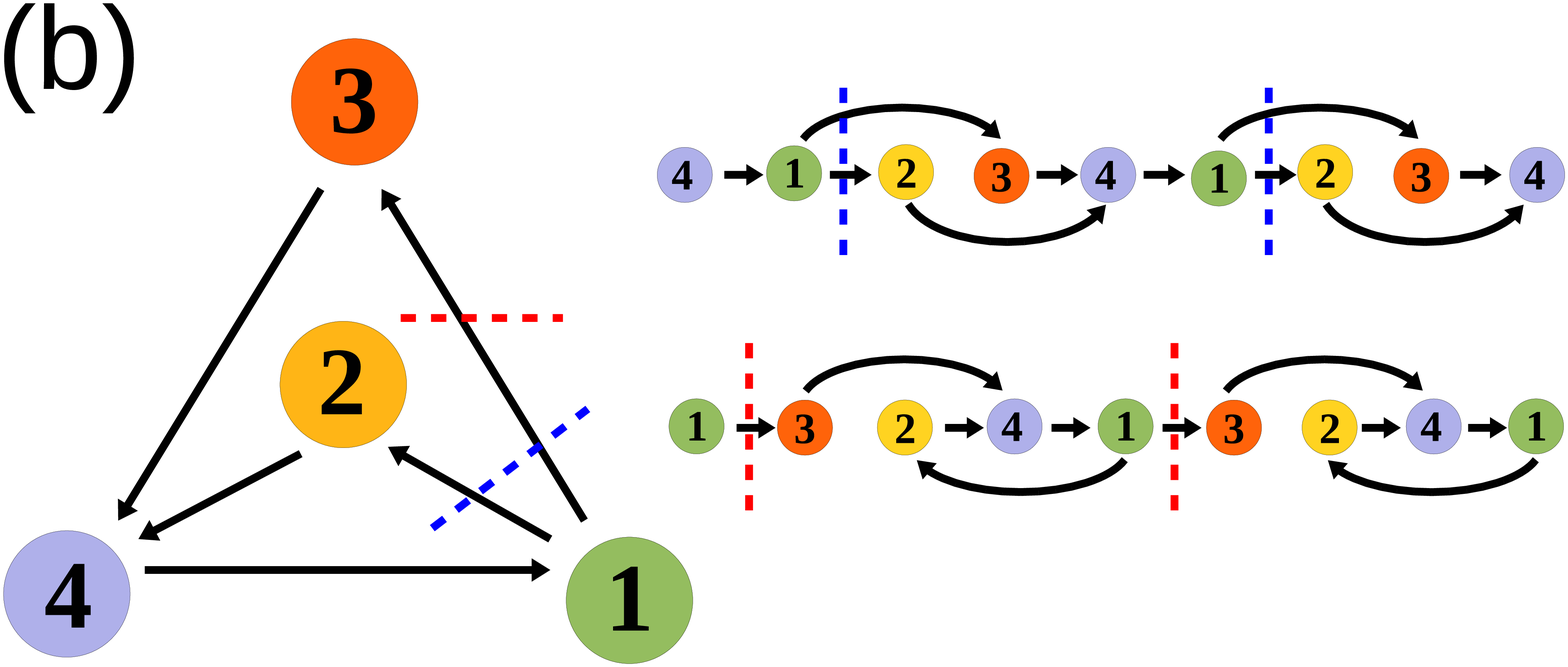}
\caption{\small (Color online) Schematics of (a) a stochastic oscillator with three metastable states in a tilted harmonic potential. The transitions are Kramers rate processes and can be perturbed through a change in the activation energy or noise strength. In this example the forward transitions are perturbed in form of a traveling wave. (b) Nonsequential stochastic oscillator with four states. The diagrams on the right show possible lifts of the Markov chain to periodic lattices and associated Poincar\'e sections. The red Poincar\'e section (lower diagram) disregards sub-threshold oscillations $1\to 2\to 4\to 1$.}
\label{Fig:OssiCartoon}
\end{figure}

The emergence of an oscillating mean field in large ensembles of noisy or nonidentical oscillators is considered the hallmark of synchronization as a collective effect \cite{Kuramoto84,Toenjes10a,WoLiVdBKa07,LSGPrager03}. 
However, the term {\em synchronization} is used differently in different contexts. It may also refer to anything from statistical correlations to phase synchronization or complete synchronization. For individual oscillators, it is appropriate to define synchronization as the adjustment of frequency due to an interaction \cite{PiRoKurths03}. Here it is in this sense that we will study synchronization for weakly driven or coupled Markov chains. If a system with frequency $\omega_0$ is driven at frequency $\omega_1$, two effects may be observed: correlation of the state of the oscillator with the phase of the driving signal and adaptation of mean frequency $\Omega$. For deterministic limit-cycle oscillators, these effects correspond to phase and frequency synchronization and for coupling stronger than a critical value, phase and frequency locking are observed. These extreme cases are not observed in the presence of noise, although the Kramers theory for weak noise predicts an exponentially small frequency difference within the synchronization regions (Sec.\ref{Sec:ContLim}). On the other hand, strong fluctuations, such as those present in noise-induced oscillations and stochastic cycles over a few states or with heterogeneous transition rates, should be considered an integral part of the system. In this case, the first-order perturbation theory can predict the correlations induced by weak external perturbation. In addition, with the second-order perturbation theory presented in this paper, it is possible to quantify the change of frequency.
\\ \\
Stochastic processes over discrete and finite sets of states have long been used as conceptual models for stochastic oscillators and proven to capture the essential mechanisms of synchronization. Discrete-state Markov rate models are typically applicable for molecular machines. Enzymes and motor proteins can be considered as molecular machines that undergo operation cycles consuming energy and performing work. Often these operation cycles are well modeled by a finite set of configurations and reaction rates, which describe the speed of transitions between the states. The transitions occur randomly whenever a substrate molecule binds, ATP is converted or through thermal activation. Molecular machines operating under non-equilibrium conditions are therefore stochastic oscillators that can be approximated by a continuous-time Markov chain model without detailed balance, i.e., with an average flow along the operation cycle (see Fig. \ref{Fig:OssiCartoon}a). The mean frequency directly equates to the productivity of a molecular machine or the turnover rate of an enzymatic reaction, which can be increased or decreased depending on the situation. This can be achieved by driving the system purposefully at a resonant frequency.
\\ \\
If the frequencies of two oscillators are sufficiently similar, mutual coupling can increase the coherence of the stochastic oscillations. This nontrivial collective effect, where the order is an emergent property of the coupled system, is an important mechanism to reduce noise in biological oscillators on meso- and micro-scales. It has, for instance, been observed in ensembles of beating heart cells \cite{TodaiCardio05}. In \cite{GoGol06} it was shown numerically that by coupling a few noisy gene regulatory circadian oscillators, phase diffusion for each oscillator decreases significantly. Another recent paper \cite{Zwick10} concludes that in some circumstances, the circadian protein phosphorylation cycle in cyanobacteria can dramatically enhance the robustness of the coexisting gene regulatory transcription-translation cycle. Both cycles were described by their respective rate processes, i.e., coupled stochastic oscillators of different types.
\\ \\
In this paper we apply the second-order perturbation theory to the master equation of a Markov rate process to derive expressions for its mean frequency $\Omega$. The adjustment of this frequency to frequency $\omega_1$ of a deterministic or stochastic driving signal is clearly identified as a parametric resonance phenomenon. For a simple jump process on a ring with discrete rotational symmetry we can also derive expressions for the phase diffusion constant in the case of external driving or mutual coupling between two stochastic oscillators. By exploiting the special structure of the second-order perturbation terms, we also present an algebraic method to maximize or minimize these terms under linear constraints on the perturbation and quadratic constraints on its power. Thus, it is possible to optimize pump flows in a diffusion process over a directed network in response to the driving frequency. Our work is closely related to the theories of stochastic resonance, ratchets and stochastic transport \cite{FreundLSG99,RomLSG10} but should be viewed in the context of synchronization.
\section{Perturbation theory for Markov rate processes}		\label{Sec:Pertheory}
A Markov rate process over a set of states $n\in\lbrace 1,\dots,L\rbrace$ is described by its matrix $\textnormal{W}$ of transition rates and a master equation
\beq		\label{Eq:MarkovChain01}
	 {\dot P}_{nn_0}(t|t_0) = \sum_{m=1}^L W_{nm}(t) P_{mn_0}(t|t_0)
\eeq
for forward-transition probabilities $P_{nn_0}(t|t_0)$ to be in state $n$ at time $t$ when the system is in state $n_0$ at a previous time $t_0$. The matrix of transition rates is $W_{nm}\ge 0$ for $m\ne n$ and $\sum_{n} W_{nm}=0$, which guarantees conservation ($\sum_n P_{nm}=1$) and nonnegativity $P_{nm}\ge 0$ of the probability at all times. Also, because of this property one eigenvalue of $\textnormal{W}(t)$ is always zero. If this eigenvalue is non-degenerate, which is the case when the Markov chain is strongly connected, all other eigenvalues have negative real part and the transition probability is asymptotically independent of the initial conditions:
\beq	\label{Eq:AsymPt01}
	P_{nn_0}(t|t_0\to -\infty) = p_n(t) .
\eeq
The perturbation theory is applicable when $\textnormal{W}(t)$ is weakly dependent on time, i.e., for $\textnormal{W}(t)=\textnormal{W}^0 + \varepsilon\textnormal{V}(t)$ and sufficiently small $\varepsilon$. From $\sum_{n} W_{nm}(t)=0$ for all $\varepsilon$ follows $\sum_{n} V_{nm}(t)=0$ at all times.
The time-dependent perturbation ansatz for the problem
\beq	\label{Eq:WeakDriving01}
	 \dot{\textnormal{P}}(t|t_0) = \left[\textnormal{W}^0 + \varepsilon \textnormal{V}(t)\right] \textnormal{P}(t|t_0)
\eeq
is to expand $\textnormal{P}(t|t_0)$ as a series in powers of $\varepsilon$
\beq	\label{Eq:PertAnsatz01}
	\textnormal{P}(t|t_0) ~=~ \textnormal{P}^{(0)}(t|t_0)  + \varepsilon \textnormal{P}^{(1)}(t|t_0) + \varepsilon^2 \textnormal{P}^{(2)}(t|t_0) + \dots	~.
\eeq
Inserting this ansatz in Eq.~(\ref{Eq:WeakDriving01}) and sorting by powers of $\varepsilon$ the dynamics of any $\textnormal{P}^{(l)}(t|t_0)$ is determined as
\beq	\label{Eq:PlODE01}
	 \dot{\textnormal{P}}^{(l)}(t|t_0) = \textnormal{W}^0 \textnormal{P}^{(l)}(t|t_0) + \textnormal{V}(t) \textnormal{P}^{(l-1)}(t|t_0),
\eeq
which, given $\textnormal{V}(t)$ and $\textnormal{P}^{(l-1)}(t|t_0)$, is an inhomogeneous, linear ordinary differential equation that can  be solved iteratively for each perturbation order. 
\\ \\
The mean frequency of a single stochastic oscillator is proportional to a sum of directed and time-averaged flows $\left\langle J_{nm}(t)\right\rangle_t = \left\langle W_{nm}(t)p_m(t) \right\rangle_t$.
In general, we write
\beq	\label{Eq:MeanOmegaGeneral01}
	\Omega = 2\pi \sum_{n,m} \Theta_{nm} \left\langle J_{nm}(t) \right\rangle_t,
\eeq
where $\Theta_{nm} \in \left\lbrace-1,0,1\right\rbrace$ defines a Poincar\'e section and the direction in which the Poincar\'e section is crossed for each transition (Fig.\ref{Fig:OssiCartoon}b).
Note that $\Omega$ depends on the choice of the Poincar\'e section, which can be arbitrary. Therefore, in general, $\Omega$ is not equal to one of the relaxation frequencies of the system. In fact, for the two state Markov model of stochastic resonance \cite{McNamWiese89}, which has the same form  as Eq.(\ref{Eq:MarkovChain01}), relaxation is not oscillatory at all, whereas $\Omega\ne 0$.
\\ \\
An alternative way to define mean frequency $\Omega$ and, in addition, phase diffusion constant $D$, of the stochastic oscillator is to lift the stochastic jump process to a periodic lattice with $W_{n+L,m+L}=W_{nm}$. It is necessary to decide whether each transition is forward or backward (Fig.\ref{Fig:OssiCartoon}b). Then the mean frequency and the phase diffusion constant are derived from the asymptotic behavior of the mean and variance of $n(t)$, for instance, with respect to distribution $P_{nn_0}(t+\tau|t)$ as
\besplit	\label{Eq:MeanMomentLimit01}
	\Omega &= \frac{2\pi}{L} \lim_{\tau\to\infty} \frac{1}{\tau} \mathbb{E}\left[n(t+\tau)-n_0\right], \\ 
	\quad D &= \left(\frac{2\pi}{L}\right)^2 \lim_{\tau\to\infty} \frac{1}{2\tau} \textnormal{Var}\left[n(t+\tau)-n_0\right],
\eesplit
where $\mathbb{E}$ and $\textnormal{Var}$ denote expected value and variance, respectively.
Because of asymptotic independence from the initial conditions, it is possible to perform a suitable average over the initial conditions $n_0$ and $t$.
The perturbation theory for this rate process over an infinite, periodic lattice requires the Bloch functions of the infinite, periodic difference operator obtained by the lift of $\textnormal{W}^{0}$. We will only derive expressions for our simplest example, which is a one-dimensional ring with discrete rotational symmetry. For this example, the Bloch waves are plain harmonics. Dimensionless quantity $c=\Omega/D$ is called the P\'eclet number and describes the coherence of a stochastic oscillator. Its value indicates the number of rotations for an ensemble of stochastic oscillators until coherence is lost.
\\ \\
Note that, in the linear order of the perturbation, all time averages depend linearly on the time average of $\textnormal{V}(t)$. Therefore, without loss of generality, we can absorb this time average into matrix $\textnormal{W}^{0}$ and assume $\left\langle\textnormal{V}(t)\right\rangle_t=0$ henceforth. Nonlinear effects are first observed in the time averages of second-order perturbation terms $\textnormal{P}^{(2)}$, $\Omega^{(2)}$, $D^{(2)}$ and $c^{(2)}$ of the transition probabilities and other quantities. We further assume that the perturbation is of the form
\beq	\label{Eq:ComplexV01}
	\textnormal{V}(t) = \textnormal{H}z(t) + \textnormal{H}^*z^*(t), 
\eeq
where complex driving signal $z(t)$ has rotational symmetry such that
\besplit	\label{Eq:zConditions01}
 &\left\langle z(t) \right\rangle_t = \left\langle z(t+\tau)z(t)\right\rangle_t = 0, \quad \textnormal{and} \\
 &\left\langle z^*(t+\tau)z(t)\right\rangle_t = e^{\Lambda \tau}	.					
\eesplit
Equations~(\ref{Eq:ComplexV01}) and (\ref{Eq:zConditions01}) include the case of harmonic driving $z(t)=\exp(-i\omega_1 t)$ where $\Lambda=i\omega_1$ and driving with a continuous or discrete stochastic signal with finite correlation time $|\textnormal{Re}[\Lambda]|^{-1}<\infty$ and relaxation frequency $\textnormal{Im}[\Lambda]=\omega_1$. Complex matrix $\textnormal{H}$ assigns relative strength and phase to the perturbation of each transition. Thus we refer to $\textnormal{H}$ as the driving protocol. 
Second-order shift $\Omega^{(2)}$ of the mean frequency is a real-valued function 
expressed as a sum of products between single entries of $\textnormal{H}^*$ and $\textnormal{H}$. A Hermitian operator corresponds to this quadratic form over the space of complex matrices, and we define the notation
\beq	\label{Eq:Hermdw01}
	\Omega^{(2)}(\textnormal{H}^*,\textnormal{H}) = 2\pi\left[\nu(\textnormal{H}^*,\textnormal{H}) + \nu^*(\textnormal{H}^*,\textnormal{H})\right],
\eeq
where the asterisk denotes the complex conjugate.
The perturbation expansion of the transition probabilities and related quantities are written in terms of the left and right eigenvectors $\mathbf{v}^{(k)}$ and $\mathbf{u}^{(k)}$ of $\textnormal{W}^0$, respectively, and corresponding eigenvalues $\lambda_k$, i.e., $\textnormal{W}^0 \mathbf{u}^{(k)}=\lambda_k \mathbf{u}^{(k)}$ and $\mathbf{v}^{(k)\dagger} \textnormal{W}^0 = \lambda^*_k \mathbf{v}^{(k)\dagger}$, where $\dagger$ denotes the complex conjugate transpose.  We assume normalization $\mathbf{v}^{(k)\dagger}\mathbf{u}^{(k')} = \delta_{kk'}$, completeness $\sum_k \mathbf{u}^{(k)}\mathbf{v}^{(k)\dagger}=\mathbb{1}_{L\times L}$ and conventions $\lambda_0=0$,  $\mathbf{p}^{(0)}=\mathbf{u}^{(0)}$ and $\mathbf{v}^{(0)}=\mathbf{1}$, where $\mathbb{1}_{L\times L}$ is the identity matrix of size $L$ and {\mbox{$\mathbf{1}=(1,\dots,1)^\top$}. In Appendix A, we derive
\besplit
\label{Eq:MainResult01}
	  \nu (\textnormal{A}^*,\textnormal{B}) = &- \sum_{nm} \Theta_{nm} \sum_{k\ne 0} u^{(k)}_m \left( 
	  \vphantom{\sum_{k'\ne 0}}
	  A^*_{nm} \frac{\mathbf{v}^{(k)\dagger}\textnormal{B}\mathbf{u}^{(0)}}{\lambda_k + \Lambda}  	  \right. \\
	  &\left. - W^0_{nm} \sum_{k'\ne 0} \frac{\mathbf{v}^{(k)\dagger}\textnormal{A}^*\mathbf{u}^{(k')}\mathbf{v}^{(k')\dagger}\textnormal{B}\mathbf{u}^{(0)}}{\left(\lambda_{k'}+\Lambda\right)\lambda_k}  \right)
\eesplit
for the bilinear form $\nu(\textnormal{A}^*,\textnormal{B})$ in Eq.~(\ref{Eq:Hermdw01}). Equations (\ref{Eq:Hermdw01}) and (\ref{Eq:MainResult01}) capture the essence of frequency adjustment toward the frequency of the driving signal (Sec. \ref{Sec:Bsp}) and also demonstrate the resonance nature of synchronization. Denominators $(\lambda_k + \Lambda)$ in the sums in Eq.~(\ref{Eq:MainResult01}) are minimized when $\textnormal{Im}[\Lambda]=-\textnormal{Im}[\lambda_k]$, i.e., when the driving frequency matches the relaxation frequency of some mode $k$. 
\section{Optimization} \label{Sec:Opt}
This section describes a corollary technique to determine the complex driving protocol $\textnormal{H}$ that optimizes the second-order perturbation response of the mean frequency.  In Sec.~\ref{Sec:Network}, we apply this technique to a Markov rate jump process on a random network.
\\ \\
Being able to express the time-averaged second-order responses as a Hermitian form in terms of eigenvalues and eigenfunctions of the unperturbed transition matrix, we can formulate an optimization problem that can be solved using linear algebra. Consider a quadratic form 
\beq	\label{Eq:Qform02}
      f(\mathbf{x})=\mathbf{x}^\dagger\textnormal{F}\mathbf{x}
\eeq
with $\mathbf{x}\in\mathbb{C}^{N}$ and a Hermitian matrix or operator $\textnormal{F}^\dagger=\textnormal{F}$. Linear constraints may be given as 
\beq	\label{Eq:LinConst02}
      \textnormal{P}^\dagger \mathbf{x} = 0,
\eeq
where the column vectors of $\textnormal{P}\in\mathbb{C}^{N\times l}$~with $l<N$ can, without loss of generality, be assumed to be orthonormal, i.e., $\textnormal{P}^\dagger\textnormal{P}=\mathbb{1}_{l\times l}$.
The nullspace of $\textnormal{P}$ is spanned by another set of orthonormal vectors given by the column vectors of matrix $\textnormal{Q}\in\mathbb{C}^{N\times (N-l)}$ such that \mbox{$\textnormal{Q}^\dagger\textnormal{Q}=\mathbb{1}_{(N-l)\times (N-l)}$} and $\textnormal{P}^\dagger\textnormal{Q}=0$.
The purpose of optimization is to maximize or minimize $f(\mathbf{x})$ subjected to the linear constraints in Eq. (\ref{Eq:LinConst02}) and the constraint on a cost function given by a positive definite matrix $\textnormal{S}$ as
\beq	\label{Eq:PowerConst02}
      \mathbf{x}^\dagger \textnormal{S} \mathbf{x} = 1.
\eeq
This problem is related to the variational problem in quantum mechanics to minimize or maximize
\beq	\label{Eq:QMvariation02}
      \Phi^\dagger\textnormal{M}\Phi \qquad\textnormal{subject to} \quad ||\Phi||^2=1
\eeq
for some Hermitian operator $\textnormal{M}$. It can be easily shown that the solution $\Phi_\textnormal{opt}$ is the normalized eigenfunction of $\textnormal{M}$ with the largest or smallest eigenvalue. To put our optimization problem in this form, we need to project into the nullspace of $\textnormal{P}$ and apply a similarity transformation that changes the constraint in Eq. (\ref{Eq:PowerConst02}) on the cost function into the normalization condition on $\Phi$. This is achieved by the following rules:
\besplit	\label{Eq:MainResult02}
      \Sigma &= \textnormal{Q}^\dagger \textnormal{S} \textnormal{Q}~,	\\
      \textnormal{M} &= \Sigma^{-1/2} \textnormal{Q}^\dagger \textnormal{F} \textnormal{Q} \Sigma^{-1/2}~,	\\
      \mathbf{x} &= \textnormal{Q} \Sigma^{-1/2} \Phi ~~.
\eesplit
Note that with positive definite $\textnormal{S}$, projected matrix $\Sigma$ is also positive definite so that $\Sigma^{-1/2}$ is well defined and Hermitian.
\\ \\
When applied to the problem of finding an optimal driving protocol, vector $\mathbf{x}\in \mathbb{C}^{L^2}$ represents complex matrices $\textnormal{H}\in\mathbb{C}^{L\times L}$. The linear constraints can be as simple as requiring $H_{n_0m_0}=0$ for impossible transitions from state $m_0$ to $n_0$ or transitions that cannot be perturbed. More subtle constraints can be assigned on the phase relations. For example, setting $H_{n_1 m_0} = e^{i\pi} H_{n_2 m_0}$ describes periodic switching in preference between two possible target states that are reached from state $m_0$, which is a very simple model for intersections with a traffic light. Optimizing the frequency and phase relations between traffic lights in a road network is a classic problem in transport theory.
\\ \\
The quadratic form $f(\mathbf{x})=\mathbf{x}^\dagger\textnormal{F}\mathbf{x}$ that is to be optimized can, for instance, be second-order frequency shift $\Omega^{(2)}(\textnormal{H}^*,\textnormal{H})$. By using the explicit expressions given in Eqs. (\ref{Eq:Hermdw01}) and (\ref{Eq:MainResult01}), we can also optimize  $\frac{\partial\Omega^{(2)}}{\partial\omega_1}$ at a fixed driving frequency $\omega_1$. A driving protocol that maximizes the response to changes in the driving frequency can be said to have good synchronizing properties.
\section{Examples}	\label{Sec:Bsp}
\subsection{Jump process on a ring} \label{Sec:RingDrive}
\begin{figure}[!t] 
%
 \includegraphics[width=4.25cm]{\picpath/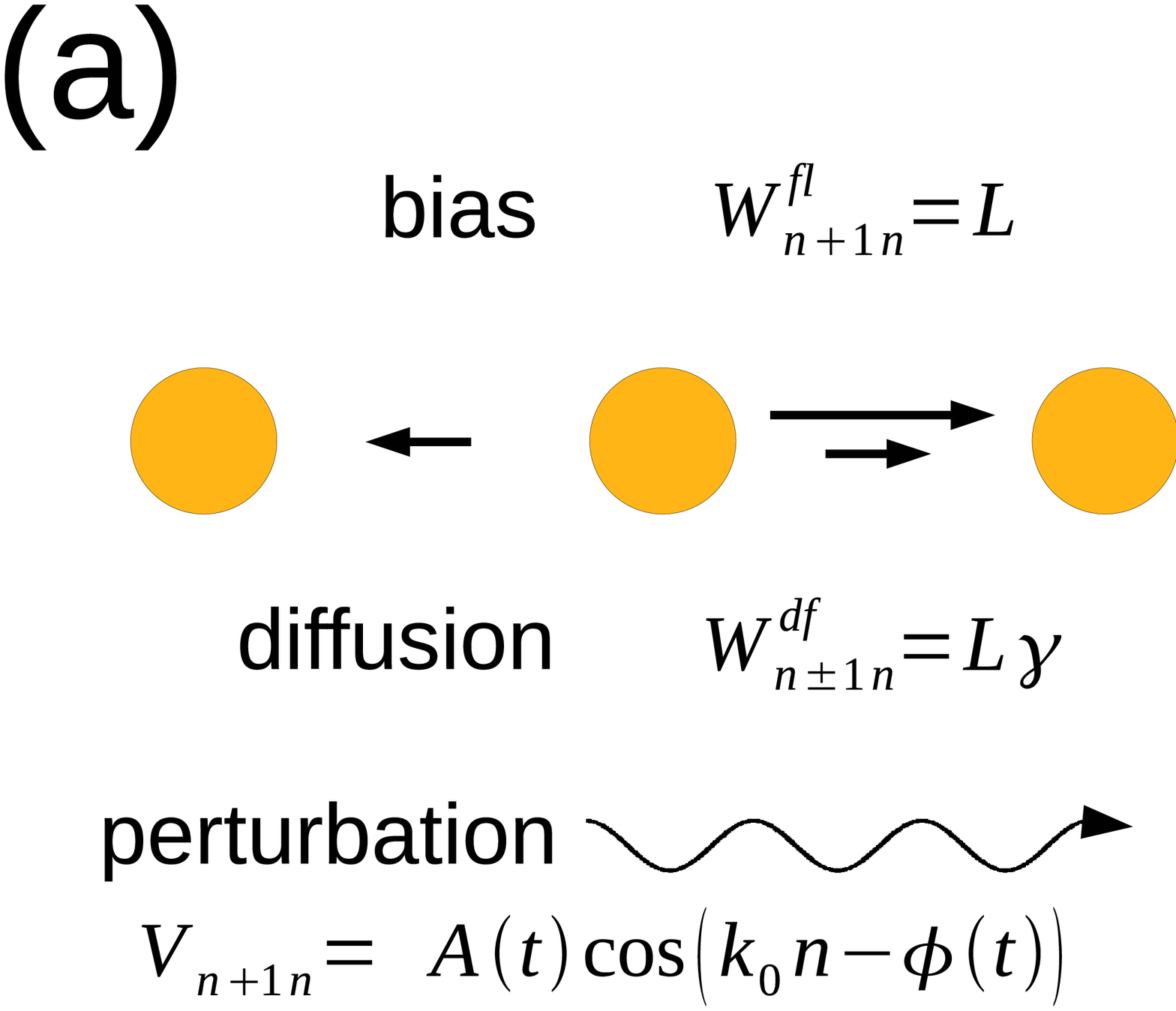} 
 \includegraphics[width=4.25cm]{\picpath/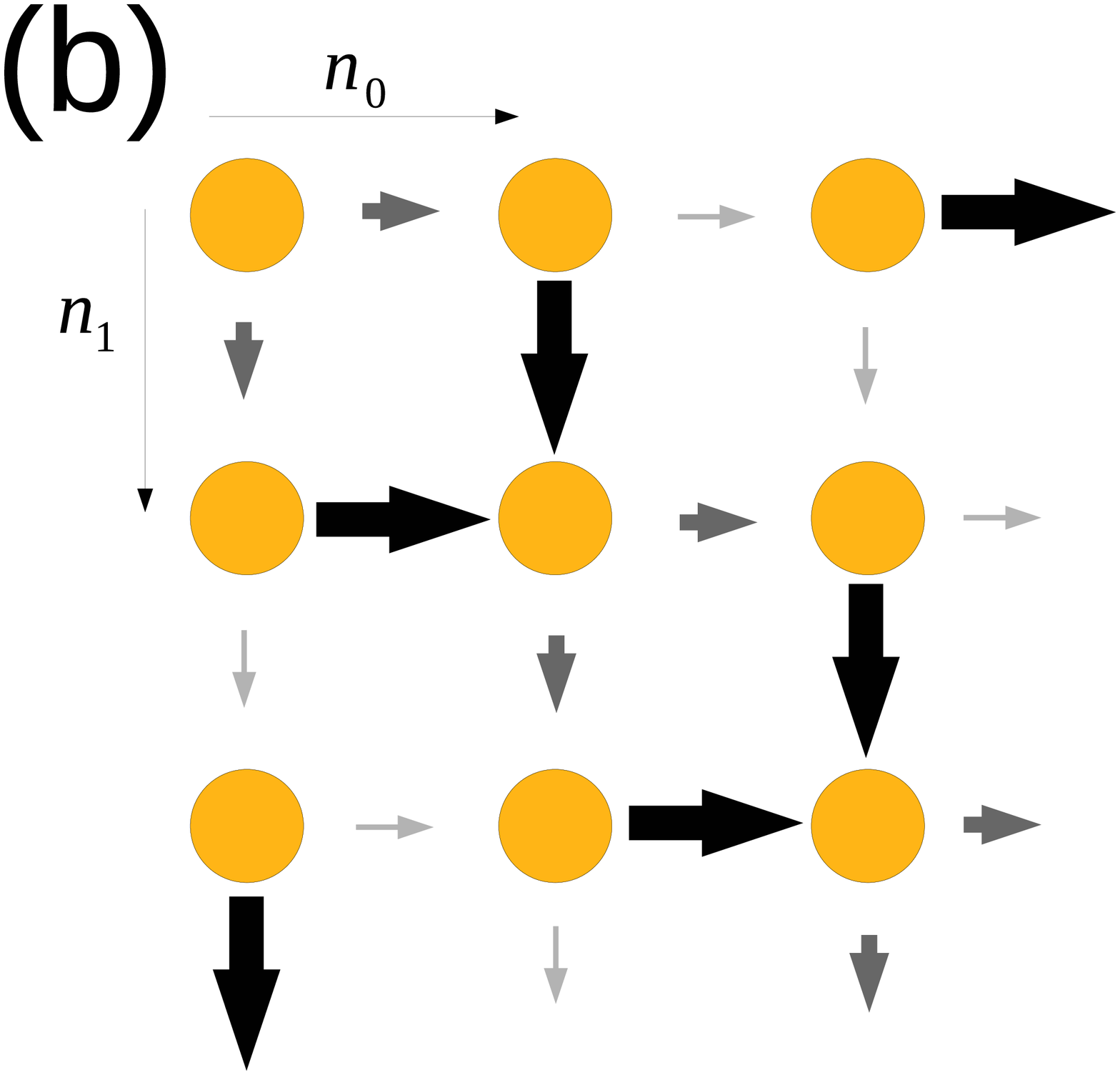}
\caption{\small (Color online) (a) Transition rates for a biased jump process on a one-dimensional periodic lattice. The transition rate is divided into a time-independent forward bias, diffusion part, and perturbation depending on time and state. (b) Transition rates for a forward jump process ($\gamma=0$) on a two-dimensional periodic lattice with attractive coupling in two directions. The speed of the transitions is indicated by different  emphasis of the arrows.}
\label{Fig:PeriodicLattices}
\end{figure}
In this section, we present the results for the mean frequency and the phase diffusion constant for a sequential change of states with discrete rotational or translational symmetry of transition rates $\textnormal{W}^{0}$.
We consider $L$ states that are visited in sequence with forward transition rates $W^0_{n+1,n} = L(1+\gamma)$ and backward transition rates $W^0_{n-1,n}=L\gamma$ (Fig.\ref{Fig:PeriodicLattices}a). We directly consider the lift of the finite state oscillator to the periodic lattice with translational symmetry $W^0_{n+1,m+1}=W^0_{nm}$. The time scale is chosen such that the mean frequency of the unperturbed oscillator is $\omega_0=2\pi$. The transition rates can be split as $\textnormal{W}^0=\textnormal{W}^{\textnormal{fl}} + \textnormal{W}^{\textnormal{df}}$ into forward jump process $\textnormal{W}^{\textnormal{fl}}$ and unbiased diffusion part $\textnormal{W}^{\textnormal{df}}$ (see Fig.\ref{Fig:PeriodicLattices}a). As discussed in Sec. \ref{Sec:ContLim}, these two rate processes give rise to the deterministic part and the diffusion of the stochastic oscillator in the continuum limit. 
\\ \\
Suppose that the perturbations are given only to the forward bias in the form of a traveling wave
\beq	\label{Eq:PertForm03}
	V_{nm} = W^{\textnormal{fl}}_{nm} A(t) \cos(k_0 m - \phi(t)),
\eeq
where phase $\phi(t)$ and amplitude $A(t)$ may be deterministic or stochastic. In terms of Eq.~(\ref{Eq:ComplexV01}), we have
\besplit
	H_{nm} 		&= W^{\textnormal{fl}}_{nm} \frac{1}{2} e^{ik_0 m}, \\
	z(t) 		&= A(t) e^{-i\phi(t)},			
\eesplit
and $z(t)$ must satisfy Eq.~(\ref{Eq:zConditions01}).
Because of the translational symmetry of the system the eigenfunctions of $\textnormal{W}^{\textnormal{fl}}$ and $\textnormal{W}^{\textnormal{df}}$ are harmonics given by $v^{(k)}_n = 2\pi u^{(k)}_n = \exp\left(ikn\right)$ with wave numbers $k\in(-\pi,\pi]$. The eigenvalues are given by
\besplit	\label{Eq:LambdaDecomp03}
      \textnormal{W}^{\textnormal{fl}} \mathbf{u}^{(k)} &= \lambda^{\textnormal{fl}}_k \mathbf{u}^{(k)} = L \left(e^{-ik}-1\right) \mathbf{u}^{(k)},	\\
      \textnormal{W}^{\textnormal{df}} \mathbf{u}^{(k)} &= \lambda^{\textnormal{df}}_k \mathbf{u}^{(k)} = L\gamma \left(e^{ik}+e^{-ik}-2\right) \mathbf{u}^{(k)},	\\
      \textnormal{W}^0 \mathbf{u}^{(k)} &= \lambda_k \mathbf{u}^{k} = (\lambda^{\textnormal{fl}}_k + \lambda^{\textnormal{df}}_k) \mathbf{u}^{(k)}.	
\eesplit
Furthermore, the action of the perturbation on an eigenmode creates two eigenmodes of wave numbers $k\pm k_0$, i.e.,
\besplit	\label{Eq:HAction03}
      \textnormal{H}\mathbf{u}^{(k)} 	&= \frac{1}{2}\lambda^{\textnormal{fl}}_{k+k_0} \mathbf{u}^{(k+k_0)} \\
      \textnormal{H}^* \mathbf{u}^{(k)} &= \frac{1}{2}\lambda^{\textnormal{fl}}_{k-k_0} \mathbf{u}^{(k-k_0)}.
\eesplit
In Appendix B, by using the perturbation expansion of the characteristic function for the random process, we determine mean frequency $\Omega$, phase diffusion constant $D$ and P\'eclet number $c=\Omega/D$ up to the second order in  perturbation strength $\varepsilon$ as
\beq	\label{Eq:ZeroOrderSol03}
	\omega_0 = 2\pi, ~~ D_0 = \frac{2\pi^2}{L}(1+2\gamma),~~c_0=\omega_0/D_0,
\eeq
\beq
\label{Eq:MeanOmegaRing03}
	\frac{\Omega}{\omega_0} = 1 - \varepsilon^2 \frac{1}{4}\left[\frac{\lambda^{\textnormal{fl}}_{k_0}}{\lambda_{k_0}+\Lambda} + \frac{\lambda^{\textnormal{fl}}_{-k_0}}{\lambda_{-k_0}+\Lambda^*}\right],
\eeq
\bmul \label{Eq:MeanPhaseDiffRing03}
	\frac{D}{D_0} = 1 + \varepsilon^2 \frac{1}{4(1+2\gamma)} \left[2\frac{\lambda^{\textnormal{fl}}_{k_0}\left(\lambda^{\textnormal{fl}}_{k_0}+i2\gamma \textnormal{Im}\left[\lambda^{\textnormal{fl}}_{k_0}\right]\right)}{(\lambda_{k_0}+\Lambda)^2} \right. \\
	\left. 
	\vphantom{\frac{\lambda^{\textnormal{fl}}_{k_0}\left(\lambda^{\textnormal{fl}}_{k_0}+i2\gamma \textnormal{Im}\left[\lambda^{\textnormal{fl}}_{k_0}\right]\right)}{(\lambda_{k_0}+\Lambda)^2}}
	- \frac{3\lambda^{\textnormal{fl}}_{k_0}+2L}{\lambda_{k_0}+\Lambda} + c.c.\right],
\emul
\bmul
\label{Eq:MeanCoherenceRing03}
	\frac{c}{c_0} = 1 + \varepsilon^2 \left[\frac{\Omega^{(2)}}{\omega_0}-\frac{D^{(2)}}{D_0}\right] \\ 
	   \shoveleft ~~~= 1 + \varepsilon^2 \frac{1}{2(1+2\gamma)}\left[
	  \vphantom{\frac{\lambda^{\textnormal{fl}}_{k_0}\left(\lambda^{\textnormal{fl}}_{k_0}+i2\gamma\textnormal{Im}\left[\lambda^{\textnormal{fl}}_{k_0}\right]\right)}{(\lambda_{k_0}+\Lambda)^2}}
	  \frac{(1-\gamma)\lambda^{\textnormal{fl}}_{k_0}+L}{\lambda_{k_0}+\Lambda} \right. \\ 
	   \left.- \frac{\lambda^{\textnormal{fl}}_{k_0}\left(\lambda^{\textnormal{fl}}_{k_0}+i2\gamma\textnormal{Im}\left[\lambda^{\textnormal{fl}}_{k_0}\right]\right)}{(\lambda_{k_0}+\Lambda)^2} + c.c.\right],
\emul
where $c.c.$ denotes the complex conjugated terms.
The six parameters of this model are length of the ring $L$ (or alternatively the time scale of the jump process on the infinite lattice), diffusion parameter $\gamma$, wave number $k_0$ of the driving traveling wave, dissipation rate $-\textnormal{Re}[\Lambda]$ and frequency $\textnormal{Im}[\Lambda]$ of the driving signal, and the perturbation strength $\varepsilon$. If we drive deterministically with $z(t)=\exp(-i\omega_1 t)$ exponent $\Lambda=i\omega_1$ is purely imaginary. The shifts of the mean frequency and the P\'eclet number for this case are shown in Figs.\ref{Fig:MainResults}a and \ref{Fig:MainResults}b, respectively. If the stochastic oscillator is coupled to the mean field from a finite ensemble or a single other stochastic oscillator, $\Lambda$ will have a negative real part that adds to the negative real part of $\lambda_{k_0}$. Thus the resonance effect decreases because the absolute values of the denominators in Eqs.~(\ref{Eq:MeanOmegaRing03})-(\ref{Eq:MeanCoherenceRing03}) increase. For instance, with $z(t)=\exp(-i k_1 n_1(t))$, depending on another unperturbed oscillator with stochastic jump process $n_1(t)$, the exponent of the autocorrelation function is simply $\Lambda=\lambda_{-k_1}$ [see Eq.~(\ref{EqB:AutoCorr02}) in Appendix B].
Driving with another stochastic oscillator of identical length, i.e., with similar stochasticity but variable mean frequency, the frequency shift is weaker, and we observe no enhancement in the P\'eclet number (see Figs. \ref{Fig:MainResults}c and \ref{Fig:MainResults}d).
\begin{figure}[!t] 
%
 \includegraphics[width=4.25cm]{\picpath/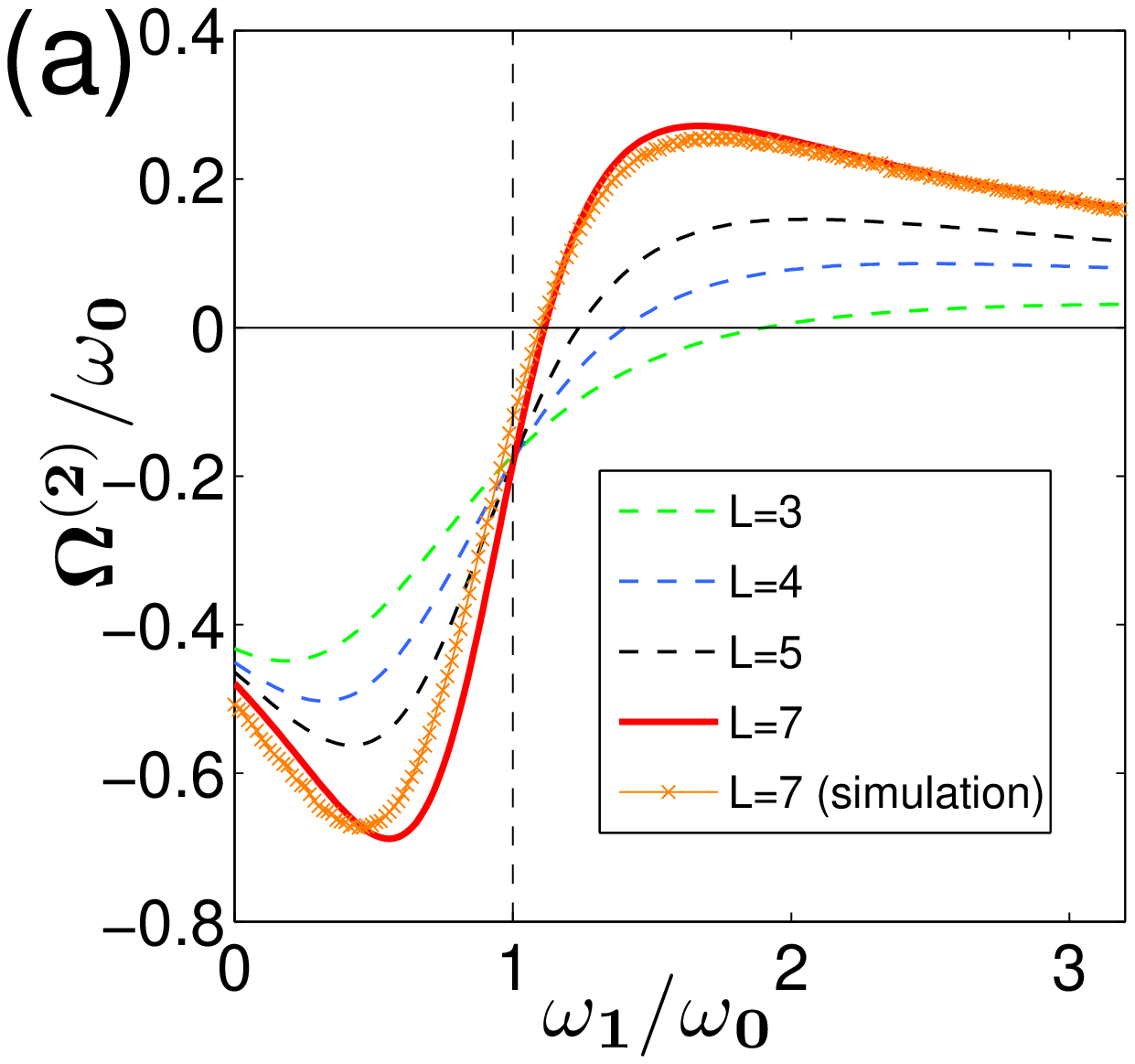} 
 \includegraphics[width=4.25cm]{\picpath/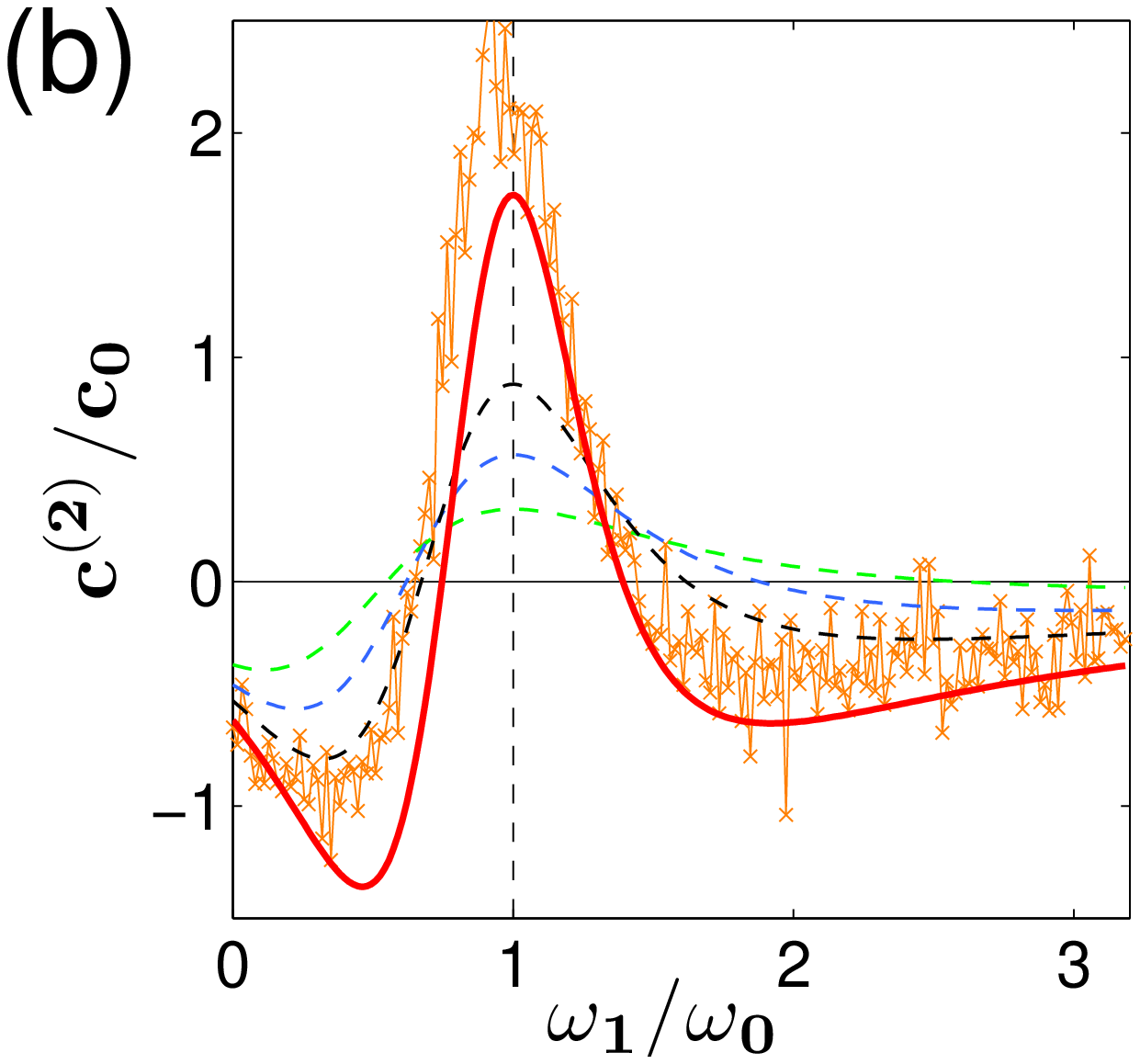}
 \includegraphics[width=4.25cm]{\picpath/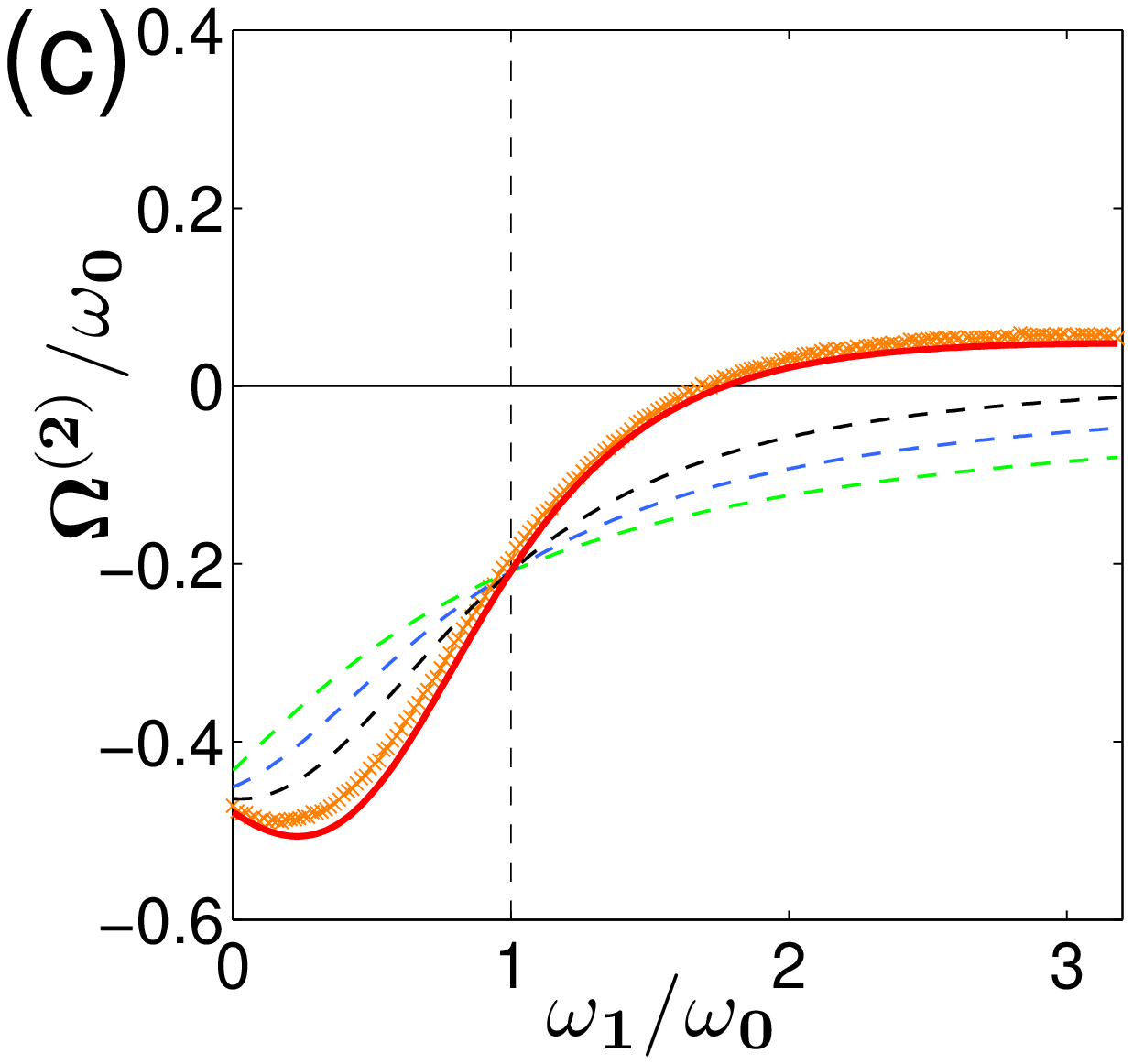} 
 \includegraphics[width=4.25cm]{\picpath/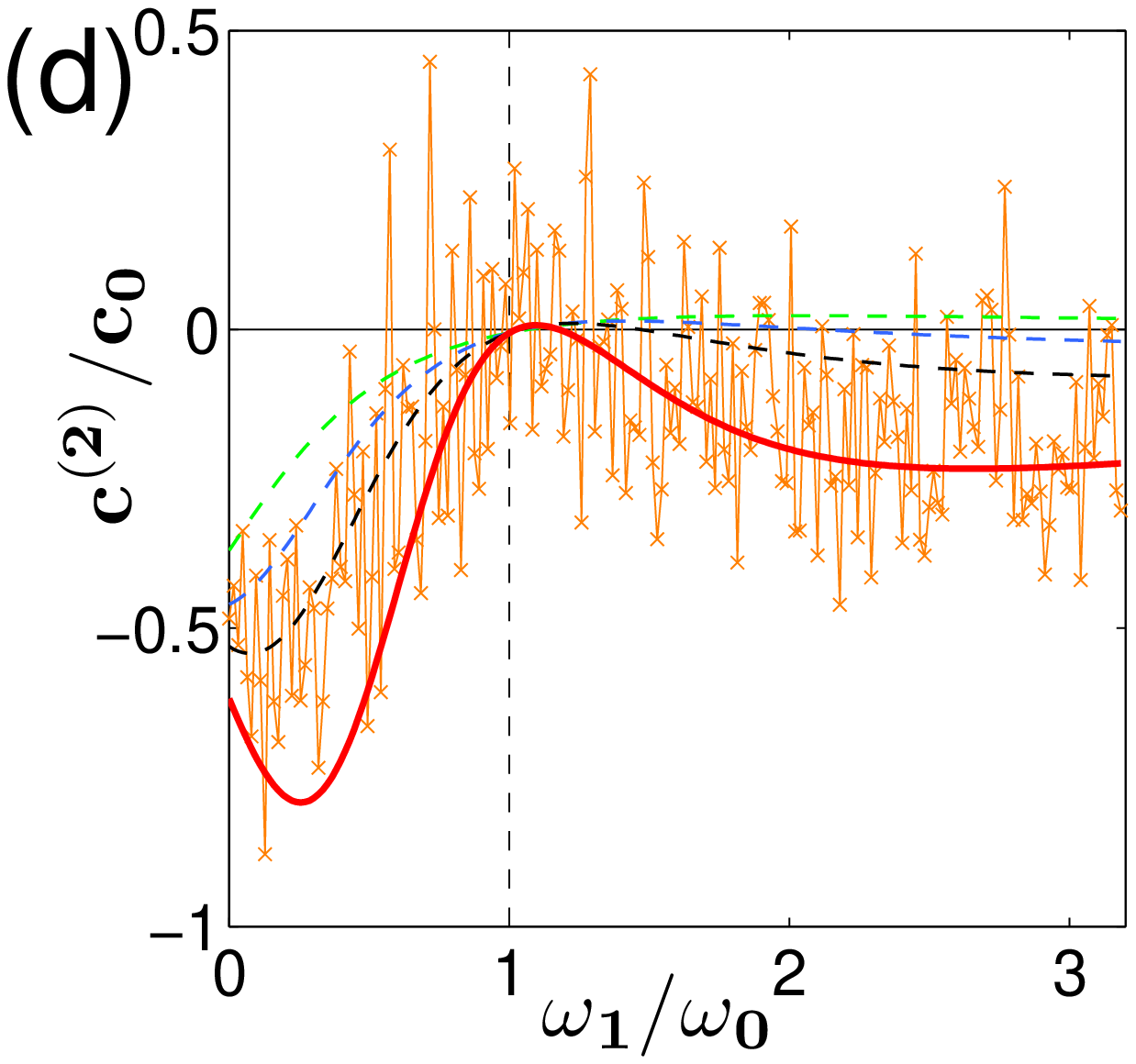}
 \includegraphics[width=4.25cm]{\picpath/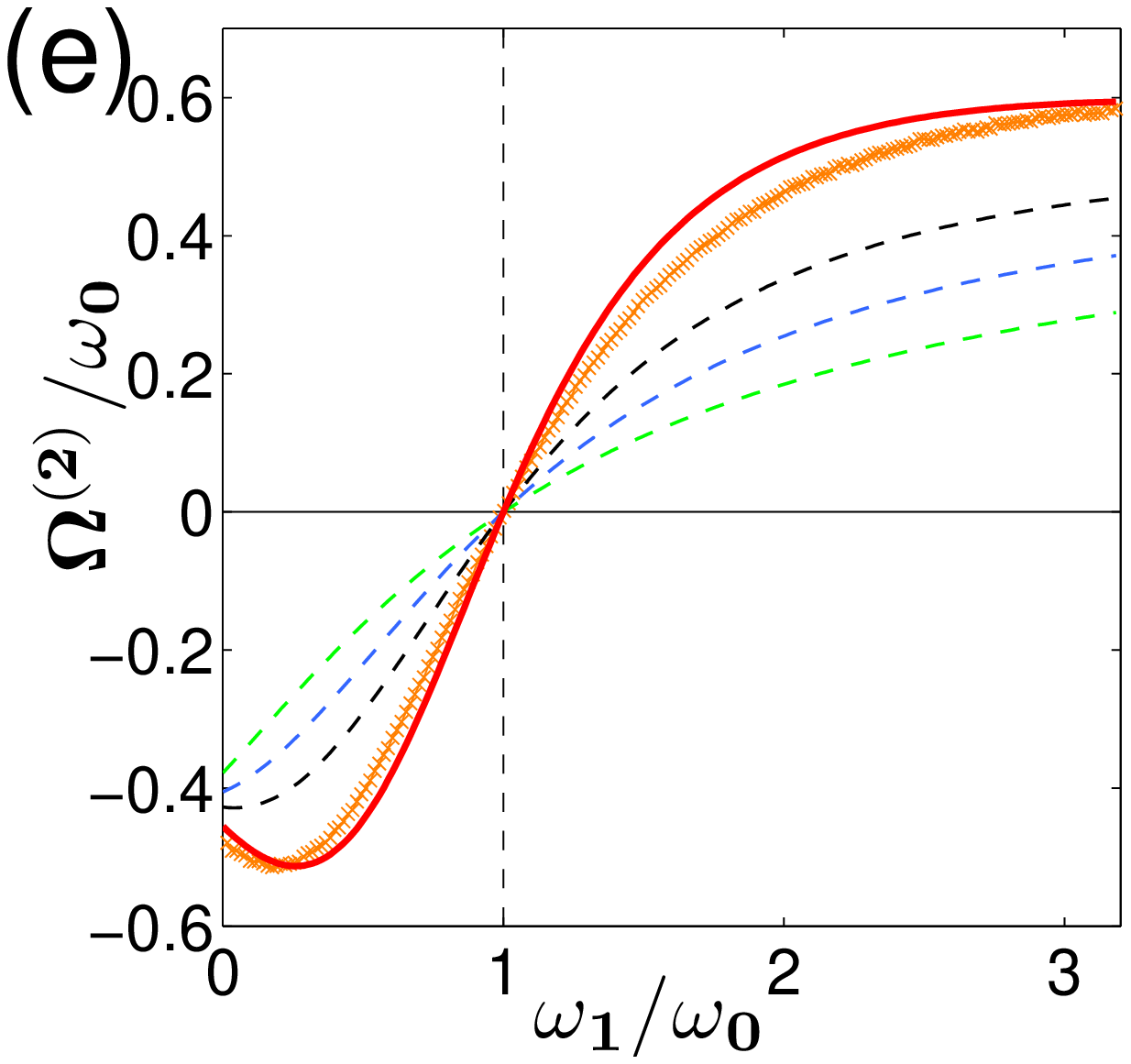} 
 \includegraphics[width=4.25cm]{\picpath/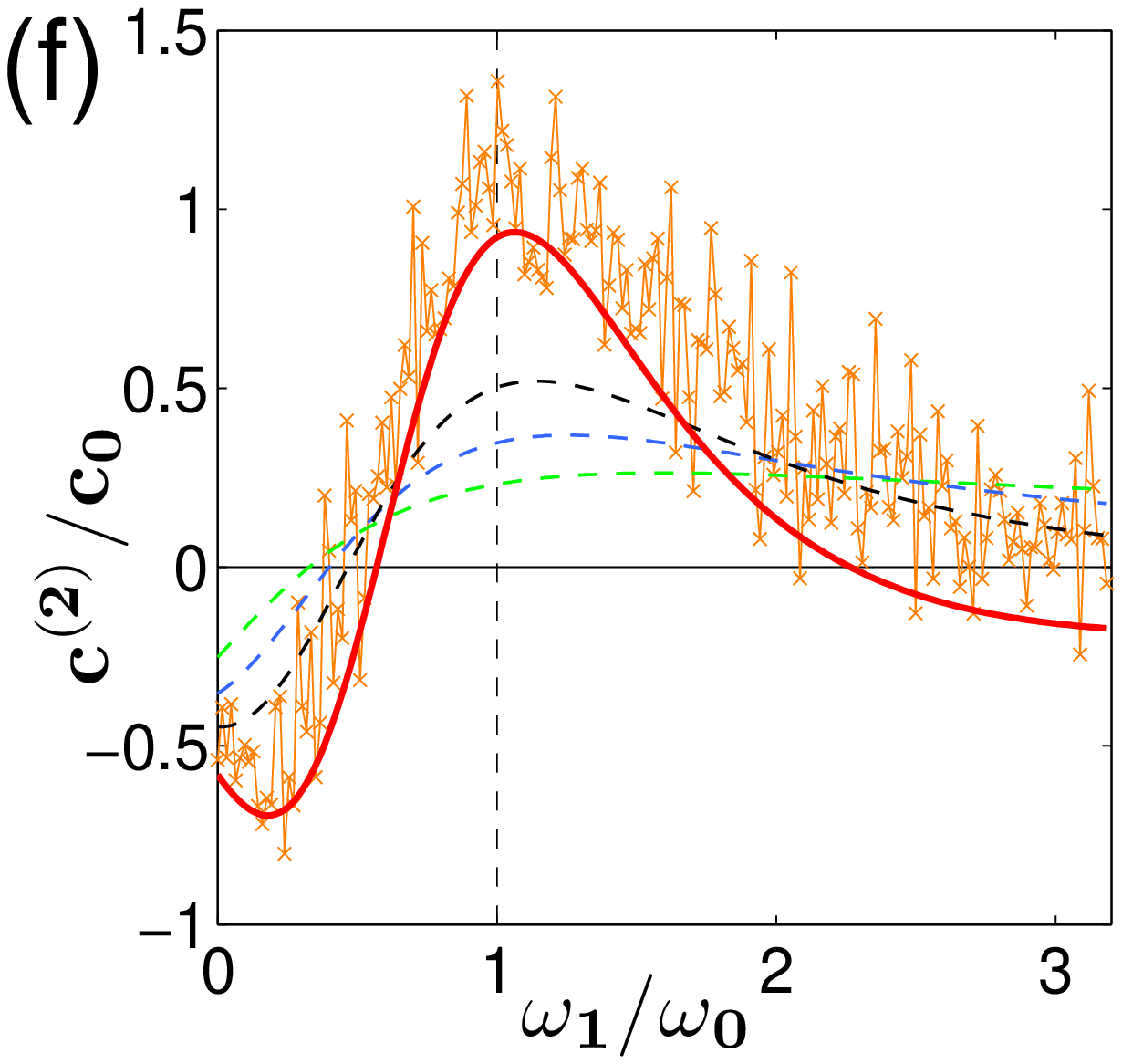}
\caption{\small (Color online) Relative frequency change (left column) and change in the P\'eclet number (right column) of a stochastic biased jump process on a ring in (a) and (b) for the case of harmonic driving of the forward transition rates with $\Lambda=i\omega_1$, in (c) and (d) for driving with another stochastic oscillator of the same length $L$ and $\Lambda^*=\lambda^{\textnormal{df}}+\lambda^{\textnormal{fl}}\omega_1/2\pi $, and in (e) and (f) for mutually coupled oscillators of same length $L$, $\mathbf{k^0}=(2\pi/L,-2\pi/L)$ and $\beta=\pi/2$. Numerical simulations were performed using a generalized Gillespie algorithm for time-dependent transition rates. Averages were taken over 1000 runs for 1000 time units each with the parameters $\gamma=0.1$, $\varepsilon=0.5$.  The inset in subfigure (a) applies to all subfigures.}
\label{Fig:MainResults}
\end{figure}
\subsection{Mutually coupled oscillators} \label{Sec:MutualCoup}
Unlike an externally driven stochastic oscillator, system $\mathbf{n}(t)=(n_0(t),n_1(t))$ of two mutually coupled stochastic oscillators is autonomous. 
As we pointed out, it is permitted to take an average over the initial conditions in Eqs.~(\ref{Eq:MeanMomentLimit01}). So far, in the time-dependent perturbation theory, averaging over time has considerably simplified the equations. Now, the perturbation does not depend on time, i.e., $z(t)=1$ in Eq.~(\ref{Eq:ComplexV01}). Instead, we average $P_{\mathbf{n_0}+\mathbf{n},\mathbf{n_0}}(\tau|0)$ over the initial conditions $\mathbf{n_0}$ (see Appendix C). We have verified our results against Monte Carlo simulations (Figs.\ref{Fig:MainResults}e and \ref{Fig:MainResults}f) of the coupled rate processes performed with the usual Gillespie algorithm for time independent transition rates.
\\ \\
We again consider the lift of the combined system $\mathbf{n}=(n_0,n_1)= n_0 \mathbf{e^0} + n_1\mathbf{e^1}$ to an infinite periodic lattice (see Fig.\ref{Fig:PeriodicLattices}b). The unperturbed transition rates may be given by isotropic diffusion part $\textnormal{W}^{\textnormal{df}}$ and forward bias $\textnormal{W}^{\textnormal{fl}}$ with different values in the directions $\mathbf{e^0}$ and $\mathbf{e^1}$ :
\besplit	\label{Eq:RateDecomp05}
	\textnormal{W}^0 &= \textnormal{W}^{\textnormal{fl}} + \textnormal{W}^{\textnormal{df}},	\\
	W^{\textnormal{df}}_{\mathbf{n}\pm\mathbf{e^0},\mathbf{n}} &= W^{\textnormal{df}}_{\mathbf{n}\pm\mathbf{e^1},\mathbf{n}} = L\gamma,	\\
	W^{\textnormal{fl}}_{\mathbf{n}+\mathbf{e^0},\mathbf{n}} = L	&,	~~W^{\textnormal{fl}}_{\mathbf{n}+\mathbf{e^1},\mathbf{n}} = \frac{\omega_1}{2\pi} L	.	\\
\eesplit
Coupling shall be given by matrix
\besplit \label{Eq:CouplingMatrix05}
	V_{\mathbf{n}+\mathbf{e^0},\mathbf{n}} &= L \cos(\mathbf{k^0}\mathbf{n}+\beta),	\\
	V_{\mathbf{n}+\mathbf{e^1},\mathbf{n}} &= \frac{\omega_1}{2\pi} L \cos(\mathbf{k^0}\mathbf{n}-\beta)	.
\eesplit
Term $\mathbf{k^0}\mathbf{n}$ denotes the inner product of $\mathbf{n}$ with wave vector $\mathbf{k}^0$. In contrast to external driving, here phase shift $\beta$ does have physical effects. For $\mathbf{k^0}=(2\pi L^{-1},-2\pi L^{-1})$ and $\beta=\pi/2$ coupling is attractive, whereas for $\beta=0$ cosine coupling inhibits synchronization. The former case is presented in Figs.\ref{Fig:MainResults}e and \ref{Fig:MainResults}f.
\\ \\
Due to symmetry, the eigenfunctions of the unperturbed system are harmonics $v_\mathbf{n}^{(\mathbf{k})} = (2\pi)^2 u_\mathbf{n}^{(\mathbf{k})} = e^{i\mathbf{k}\mathbf{n}}$. We define
\besplit	\label{Eq:LambdaDecomp05}
	\lambda^{\textnormal{df}}_{\mathbf{k}} &= \lambda^{\textnormal{df}}_{k_0} + \lambda^{\textnormal{df}}_{k_1}, \\ 
	\lambda^{\textnormal{fl}}_{\mathbf{k}}(\beta) &= \lambda^{\textnormal{fl}}_{k_0} + \frac{\omega_1}{2\pi}\lambda^{\textnormal{fl}}_{k_1} e^{-i2\beta},	\\
	\lambda_{\mathbf{k}} &= \lambda^{\textnormal{fl}}_{\mathbf{k}}(0) + \lambda^{\textnormal{df}}_{\mathbf{k}},
\eesplit
where $\lambda^{\textnormal{fl}}_k$ and $\lambda^{\textnormal{df}}_k$ are given by Eq.~(\ref{Eq:LambdaDecomp03}) and $\mathbf{k}=(k_0,k_1)$.
In Appendix C we determine the frequency, phase diffusion constant and P\'eclet number of the first oscillator to the second-order in perturbation strength $\varepsilon$ as
\beq
\label{Eq:MeanOmegaRing05}
	\frac{\Omega}{\omega_0} = \left( 1 - \varepsilon^2 \frac{1}{4}\left[\frac{\lambda^{\textnormal{fl}}_{\mathbf{k^0}}(-\beta)}{\lambda_{\mathbf{k^0}}} 
	                                                                   + \frac{\lambda^{\textnormal{fl}}_{-\mathbf{k^0}}(\beta)}{\lambda_{-\mathbf{k^0}}}\right]\right),
\eeq
\bmul \label{Eq:MeanPhaseDiffRing05}
	\frac{D}{D_0} = 1 + \varepsilon^2 \frac{1}{4(1+2\gamma)}  \left[2 \frac{\lambda^{\textnormal{fl}}_{\mathbf{k^0}}(-\beta)\left(\lambda^{\textnormal{fl}}_{k^0_0}+i 2\gamma \textnormal{Im}\left[\lambda^{\textnormal{fl}}_{k^0_0}\right]\right)}{\lambda^2_{\mathbf{k^0}}} \right. \\
	\left. 
	\vphantom{\frac{\lambda^{\textnormal{fl}}_{\mathbf{k^0}}(-\beta)\left(\lambda^{\textnormal{fl}}_{k^0_0}+i 2\gamma \textnormal{Im}\left[\lambda^{\textnormal{fl}}_{k^0_0}\right]\right)}{\lambda^2_{\mathbf{k^0}}}}
	- \frac{2 (\lambda^{\textnormal{fl}}_{k^0_0}+L)+\lambda^{\textnormal{fl}}_{\mathbf{k^0}}(-\beta)}{\lambda_{\mathbf{k^0}}} + c.c 
	\right],
\emul
\bmul
\label{Eq:MeanCoherenceRing05}
	\frac{c}{c_0} = 1 + \varepsilon^2 \frac{1}{2(1+2\gamma)}\left[ 
	\vphantom{\frac{\lambda^{\textnormal{fl}}_{\mathbf{k^0}}(-\beta)\left(\lambda^{\textnormal{fl}}_{k^0_0}+i 2\gamma \textnormal{Im}\left[\lambda^{\textnormal{fl}}_{k^0_0}\right]\right)}{\lambda^2_{\mathbf{k^0}}}}
	\frac{\lambda^{\textnormal{fl}}_{k_0^0} - \gamma \lambda^{\textnormal{fl}}_\mathbf{k^0}(-\beta) + L}{\lambda_\mathbf{k^0}}\right. \\
	\left.
	\vphantom{}
		- \frac{\lambda^{\textnormal{fl}}_{\mathbf{k^0}}(-\beta)\left(\lambda^{\textnormal{fl}}_{k^0_0}+i 2\gamma \textnormal{Im}\left[\lambda^{\textnormal{fl}}_{k^0_0}\right]\right)}{\lambda^2_{\mathbf{k^0}}} + c.c.
	\right].
\emul
In contrast to the driven system, frequency $\omega_1$ of the second oscillator enters both the denominator and numerator of the expressions via $\lambda_\mathbf{k^0}$ and $\lambda^{\textnormal{fl}}_\mathbf{k^0}(-\beta)$, respectively. As a result, the second-order perturbation terms do not vanish in the limit $\omega_1\to\infty$, since the phases of the fast and the slow oscillator remain correlated. Another remarkable difference is the possibility of an increase in coherence for two mutually coupled, identical or weakly nonidentical stochastic oscillators (Fig.\ref{Fig:MainResults}f).
\\ \\
While the time-dependent perturbation theory can be used to approximate the frequency and the phase diffusion constant under driving by another stochastic oscillator, the transition rates in the combined system $(n_0(t),n_1(t))$ are not time dependent, irrespective of mutual or unidirectional coupling. The mean frequency in the combined autonomous system can, in principle, be obtained nonperturbatively from the stationary probability distribution, i.e., the explicit zero eigenvector of the combined time-independent matrix of transition rates. In \cite{FreundNeiLSG00} this was performed for a master-slave pair of two-state oscillators, and frequency locking was observed for strong coupling. The phase diffusion constant cannot be obtained in such a simple way. In \cite{PragerLSG05} a method has been recently found to determine the phase diffusion constant in a periodically driven system from the cyclostationary solution of an extended, periodically driven, linear ordinary differential equation. In that case, the lift to the periodic lattice and the generating function approach is not necessary. It would be worth investigating whether this method can be generalized to a larger number of states, nonsequential transition networks, or stochastic driving.
\subsection{Continuum limit} \label{Sec:ContLim}
The purpose of this section is to illustrate that our theory is accurate in the limit of weak coupling and strong noise and complements the Kramers rate theory in the other asymptotic regime of strong coupling and weak noise.
\\ \\
In the limit $L\to\infty$, the biased jump process on a ring network with harmonic driving corresponds to a continuous, weakly perturbed limit-cycle oscillator in the Kuramoto phase approximation
\beq	\label{Eq:ContLimDyn04}
      \dot\vartheta = 2\pi \left(1 + \varepsilon \cos(\vartheta-\omega_1 t)\right) + \sqrt{2D_0} \xi(t)
\eeq
with delta correlated white noise $\left\langle\xi(t)\xi(t')\right\rangle = \delta(t-t')$. 
To obtain a finite phase diffusion constant, $\gamma/L$ must be kept constant such that, in the limit $L\to\infty$,
\beq \label{Eq:ContLimD004}
      \frac{4\pi^2\gamma}{L} \to D_0.
\eeq
The second-order perturbation terms for the frequency shift, the phase diffusion constant and the P\'eclet number in the continuum limit are given in Appendix B.
However, exact nonperturbative expressions for these quantities can be found. Substituting $\varphi=\vartheta-\omega_1 t$ and $\Delta=(\omega_1-2\pi)$ in Eq.~(\ref{Eq:ContLimDyn04}), we obtain the stochastic Adler equation
\beq \label{Eq:ContLimAdler04}
      \dot \varphi = -\Delta + 2\pi\varepsilon \cos\varphi + \sqrt{2D_0} \xi(t) ,
\eeq
for which the mean frequency and the phase diffusion constant are known in terms of special integrals \cite{VdBHaenggi01,SchwaPiko10}. Here, we will use the continuum limit to compare the different regimes in which our perturbation theory and the Kramers theory are valid. Noting that we use a time scale such that $\omega_0=2\pi$, we introduce dimensionless parameters $x=\Delta/D_0$ and $y=2\pi\varepsilon/D_0$. The continuum limit of the second-order perturbation frequency shift in Eq. (\ref{Eq:MeanOmegaRing03}) is given in Appendix B [Eq.~(\ref{EqB:ContLimOhm})]. In terms of $x$ and $y$, Eq.~(\ref{EqB:ContLimOhm}) can be rewritten as
\beq	\label{Eq:ContLimOhm04}
	 \frac{\Omega-\omega_0}{D_0} \approx \frac{y^2}{2} ~\frac{x}{1 + x^2}~~.
\eeq
\begin{figure}[!t] 
%
 \includegraphics[width=7cm]{\picpath/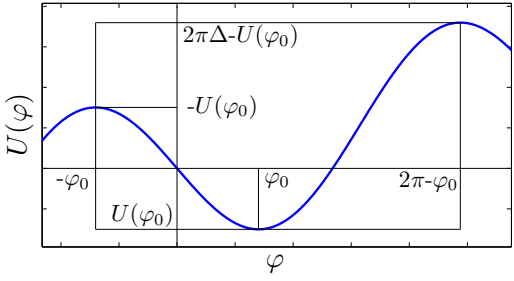} 
\caption{\small (Color online) Plot of tilted potential (bold, blue line) in the stochastic Adler equation (\ref{Eq:ContLimPot04}). The local minimum at $\varphi_0$ also determines the two adjacent local maxima at $-\varphi_0$ and $2\pi-\varphi_0$. The potential difference over the left barrier is $-2U(\varphi_0)$ and $2\pi\Delta-2U(\varphi_0)$ over the right barrier.}
\label{Fig:TiltedPotential}
\end{figure}
The derivative of this quantity with respect to $x$ is maximal at $x=0$ with
\beq	\label{Eq:ContLimOhmprime04}
	\left.\partial_{x} \frac{\Omega-\omega_0}{D_0} \right|_{x=0} = \frac{y^2}{2}~~.
\eeq
A value of this derivative larger than unity is unphysical, because then the frequency response is stronger than the change in the driving frequency. Therefore, we say $y=\sqrt{2}$ is an upper bound for the region in which our perturbation theory is valid. On the other hand, the Kramers approximation gives the rate of phase slips in the Arnold tongue region for small noise strength $D_0$ or strong coupling. Equation (\ref{Eq:ContLimAdler04}) can be written as a gradient system
\besplit	\label{Eq:ContLimPot04}
	\dot \varphi &= - U'(\varphi) + \sqrt{2D_0}\xi(t), \\
	U(\varphi) &= \left(\Delta\cdot\varphi - 2\pi\varepsilon\sin\varphi\right).
\eesplit
The tilted sine potential is shown in Fig.\ref{Fig:TiltedPotential}. It has a local minimum at $\varphi_0=\arccos(x/y)\ge 0$ and two adjacent maxima at $\varphi_1=-\varphi_0$ and $\varphi_2=2\pi-\varphi_0$. With $U_{\min}=U(\varphi_0)$ and $U_{\max}=U(\varphi_1)$ or $U_{\max}=U(\varphi_2)$, the Kramers theory requires $(U_{\max}-U_{\min})/D_0 \gg 1$.
We can parameterize the condition $(U_{\max}-U_{\min})/D_0 = 1$ by $0\le\varphi_0\le\pi/2$ and obtain
\beq	\label{Eq:CondLimKramersCond04}
      y = \frac{1}{2(\sin\varphi_0-\varphi_0\cos\varphi_0)},\qquad	x = \pm~y\cos\varphi_0 .
\eeq
Kramers rate $r_K$ over a potential barrier is given \cite{Risken89} by
\beq
	r_K = \frac{1}{2\pi} \sqrt{U''_{\min}|U''_{\max}|} e^{(U_{\min}-U_{\max})/D_0}.
\eeq
From $\Omega = \langle\dot\theta\rangle_t = \omega_1 + \langle\dot\varphi\rangle_t$ and by subtracting the backward jump rates from the forward jump rates, we obtain
\beq
	\frac{\Omega-\omega_0}{D_0} \approx x - y \sin(\varphi_0) ~e^{2(x\varphi_0-y\sin\varphi_0)} \left(1-e^{-2\pi x}\right).
\eeq
\begin{figure}[!t] 
%
 \includegraphics[height=3.7cm]{\picpath/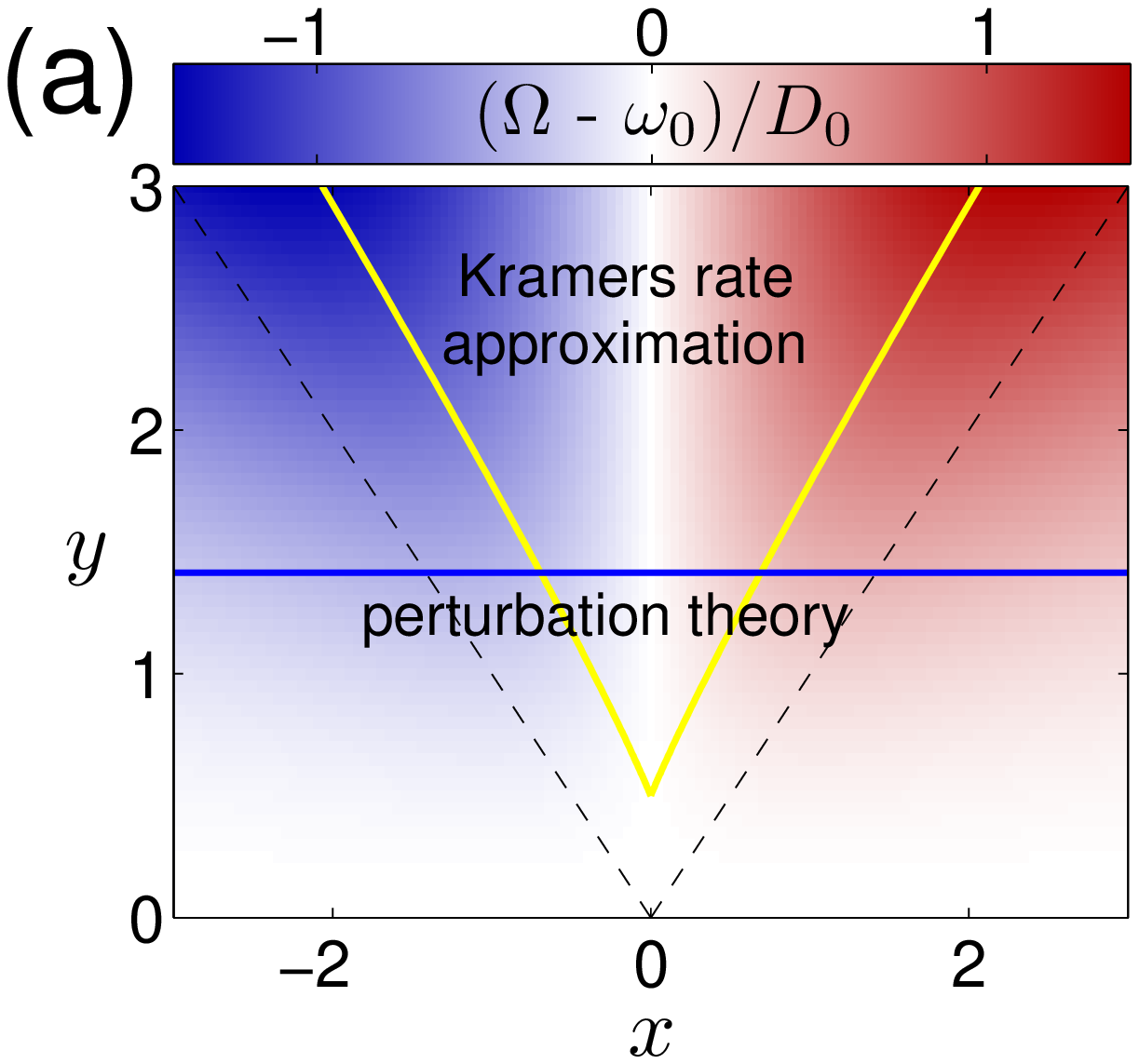} 
 \includegraphics[height=3.7cm]{\picpath/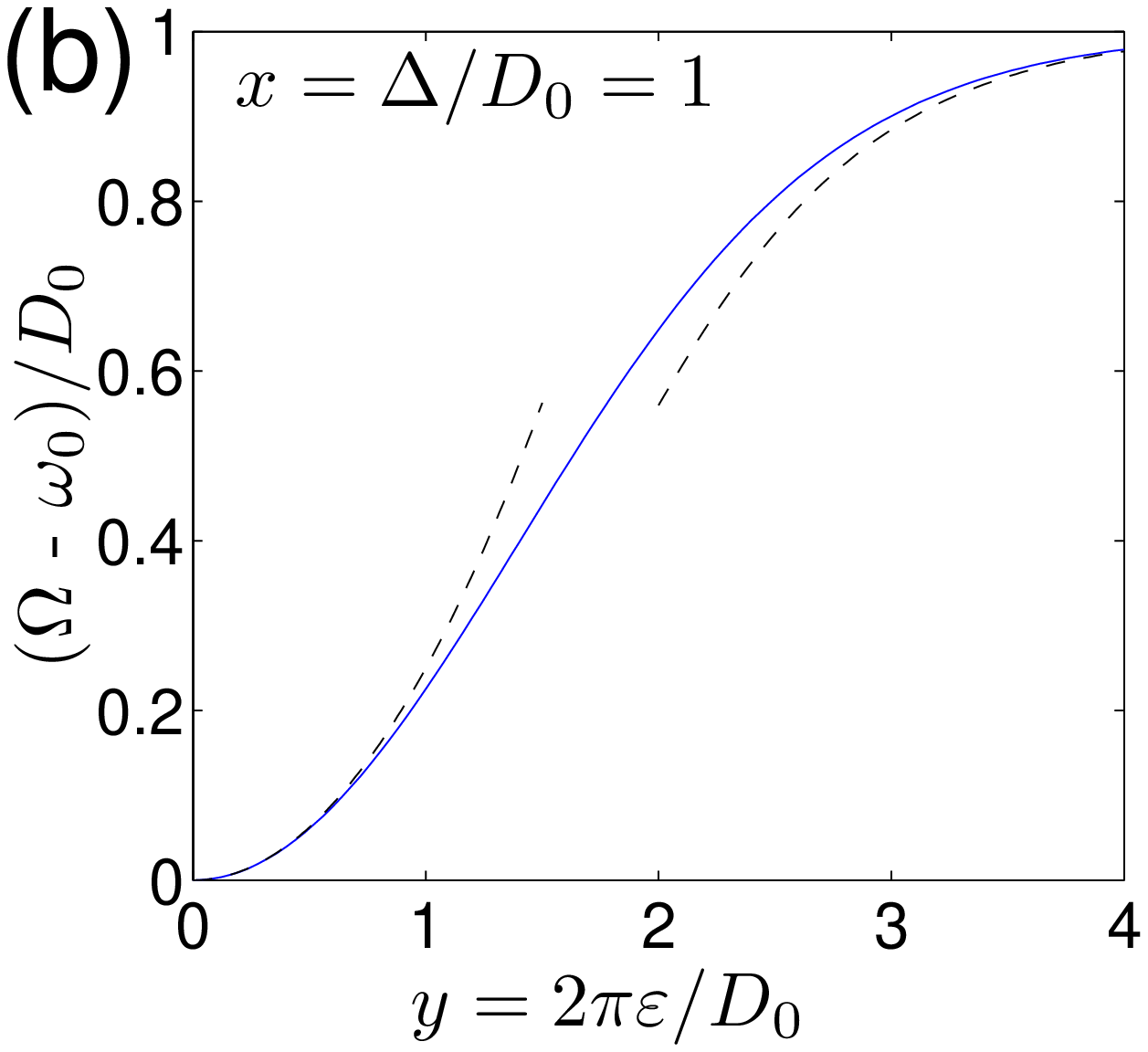} 
\caption{\small (Color online) Rescaled frequency shift $(\Omega-\omega_0)/D_0$ of the periodically driven drift- and diffusion process on a circle (a) as a function of $x=\Delta/D_0$ and $y=2\pi\varepsilon/D_0$. Values $x<0$ and $x>0$ correspond to negative (blue) and positive (red) frequency shifts, respectively (color bar aligned on top). The dashed lines mark the border of the Arnold tongue region of phase locking in the limit $D_0\to 0$. The Kramers rate theory is valid between the yellow (light, solid) curves that indicate the condition $(U_{\max}-U_{\min})/D_0=1$, whereas below the blue (dark, solid) line for $y<\sqrt{2}$, our perturbation theory will give physically sensible results for all frequency differences, in particular, at resonance point $\Delta=0$. Above the blue (dark, solid) line, the perturbation theory may be still give good results sufficiently far away from the resonance point. (b) Frequency shift as a function of $y$ with $x=1.0$ fixed (solid curve). The two dashed curves indicate the second-order perturbation approximation for low values of $y$ and the Kramers rate approximation for high $y$. The exact frequency shift was determined by numerical evaluation of the integrals in \cite{VdBHaenggi01}.}
\label{Fig:Kramers_vs_Pert}
\end{figure}
Figure \ref{Fig:Kramers_vs_Pert} compares the regions in the $(x,y)$ parameter space  where the Kramers approximation holds with that where our second-order perturbation theory is valid at the resonance point. Small noise or large coupling strengths will result in strongly nonlinear behavior near the resonant frequency and render the perturbation theory invalid. In contrast, the regime of the Kramers approximation can always be reached by increasing the coupling strength, or if $|x/y|<1$, by reducing the noise strength. Figure \ref{Fig:Kramers_vs_Pert}b shows the rescaled frequency shift at $x=\Delta/D_0=1$ as a function of $y=2\pi\varepsilon/D_0$. At low coupling strengths $y<\sqrt{2}$, the second-order perturbation theory approximates the quadratic behavior of the frequency shift, whereas at high coupling strengths, the Kramers rate theory describes the exponential deviation from frequency locking. Therefore, it is appropriate to consider the adaptation of frequency in our theory as a weak form of synchronization.
\subsection{Ring with random shortcuts}	\label{Sec:Network}
\begin{figure}[!t] 
%
 \includegraphics[width=4.25cm]{\picpath/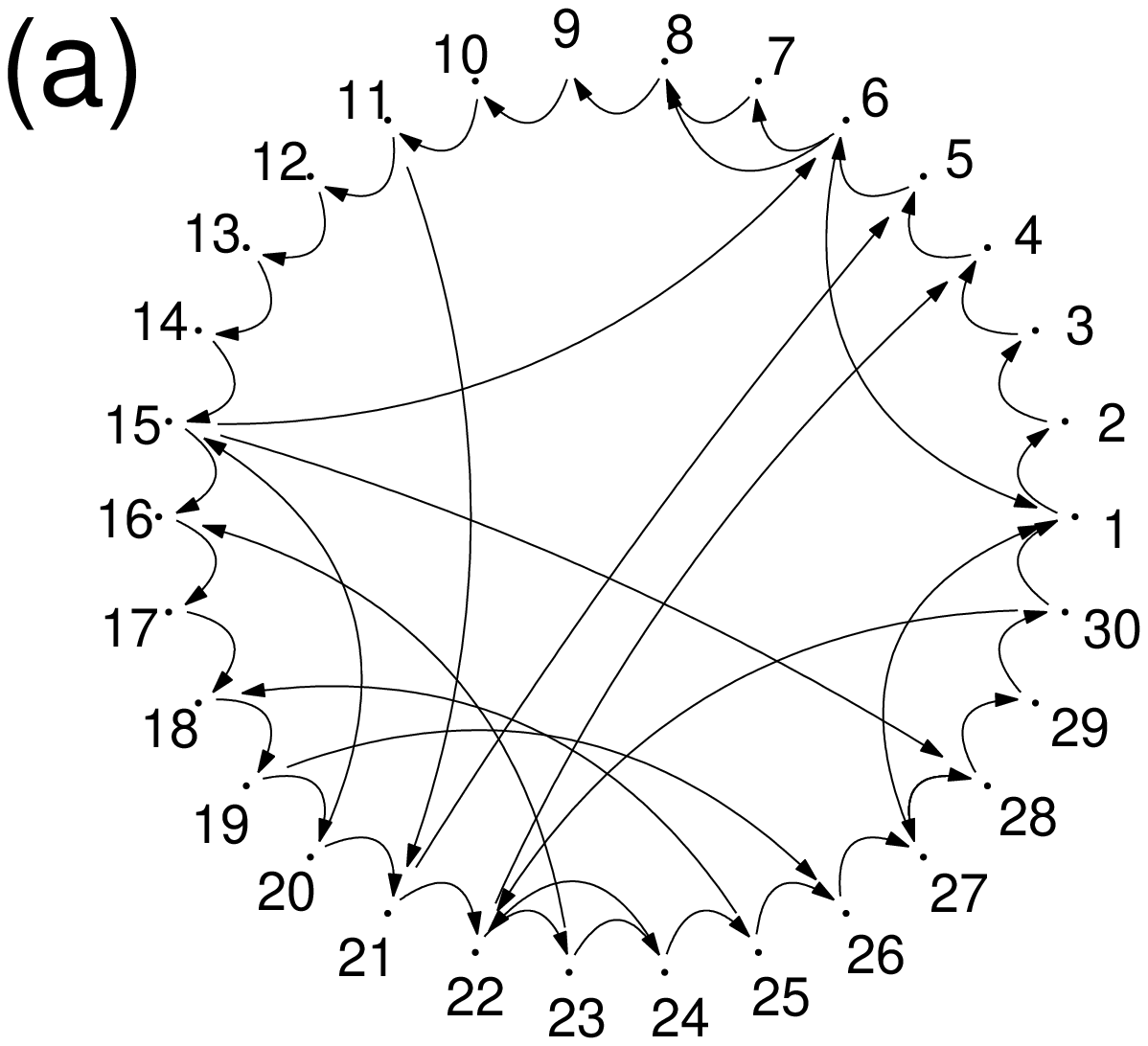} 
 \includegraphics[width=4.25cm]{\picpath/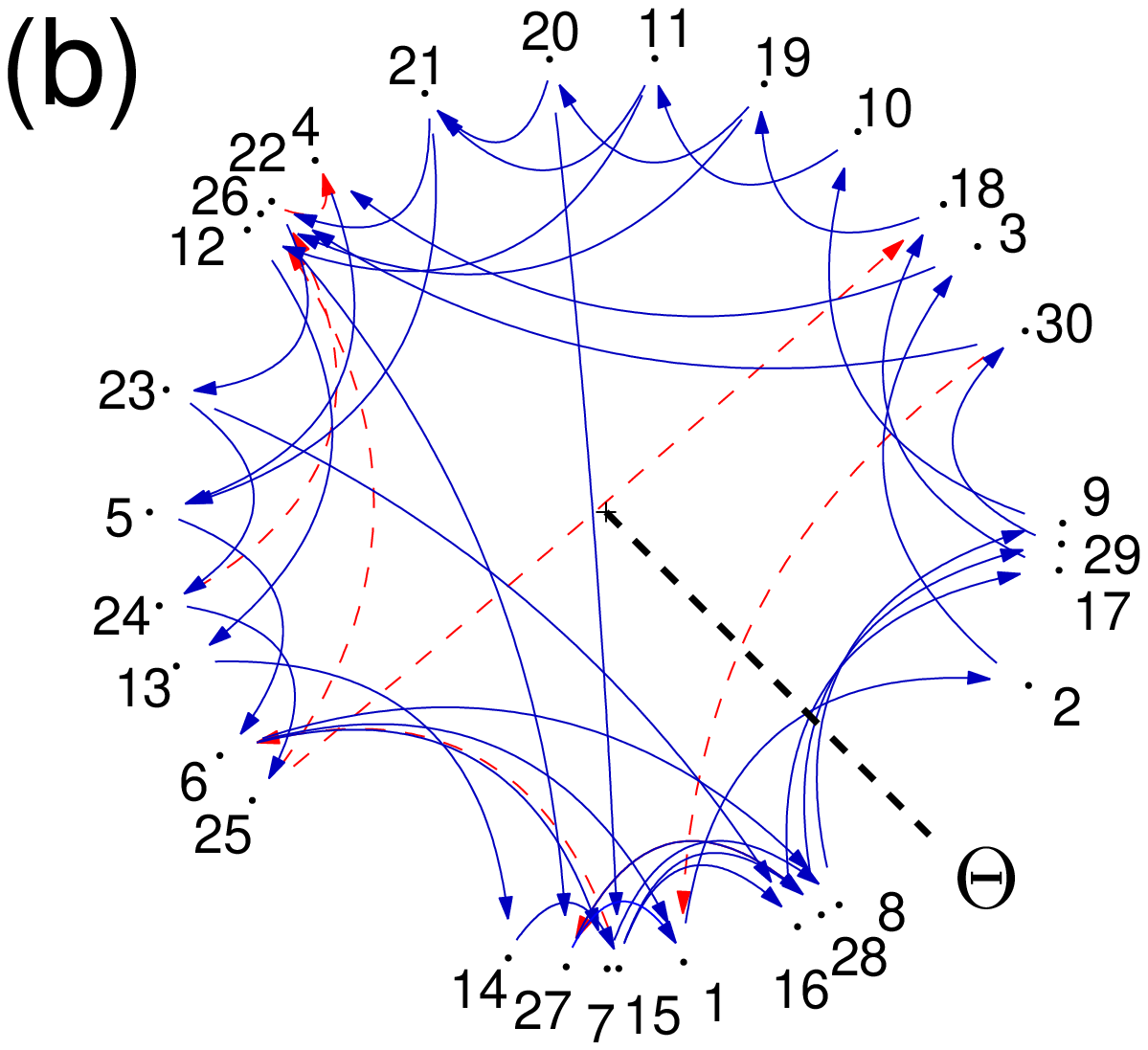}
 \includegraphics[height=3.92cm]{\picpath/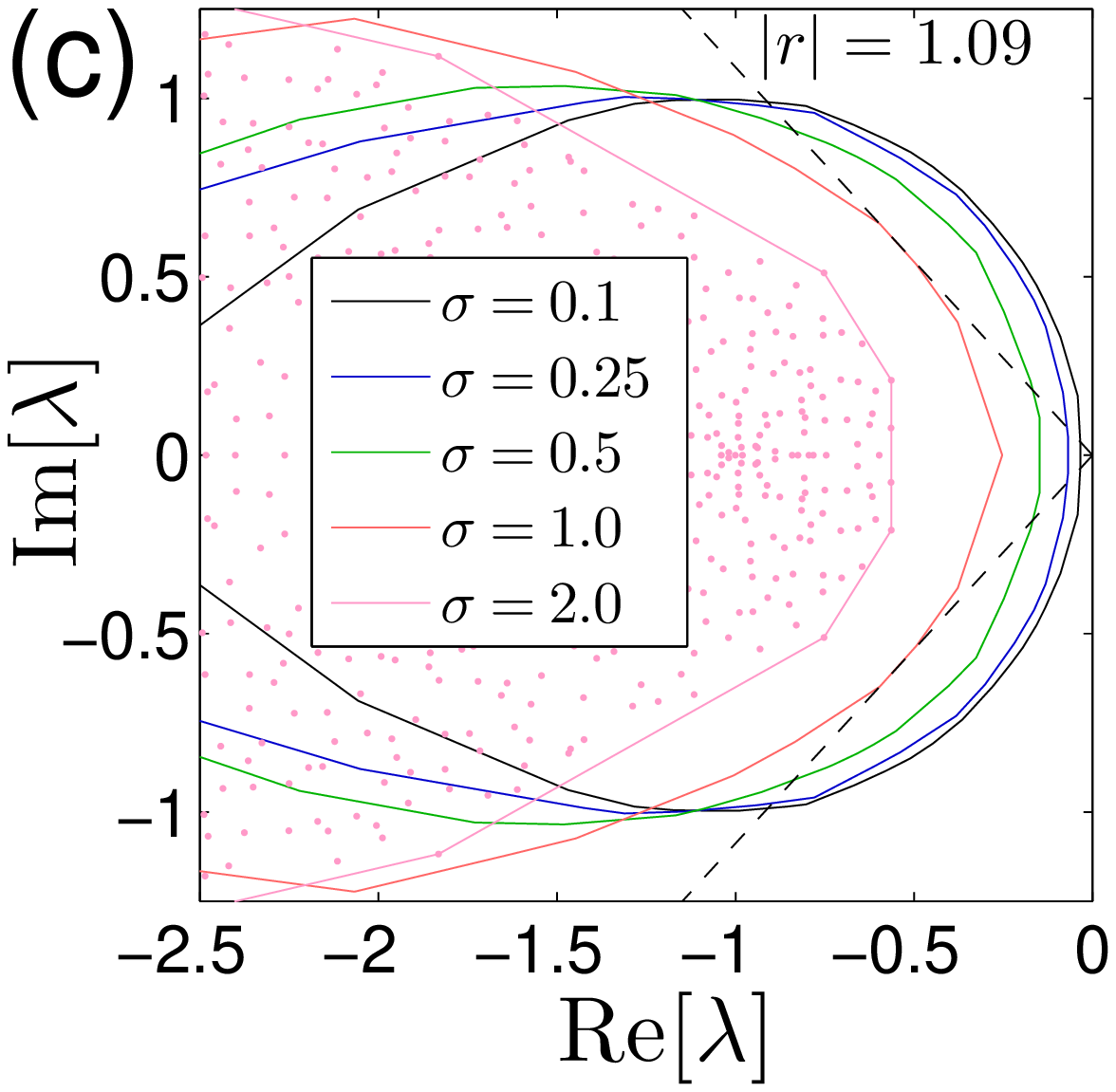}
 \includegraphics[height=3.92cm]{\picpath/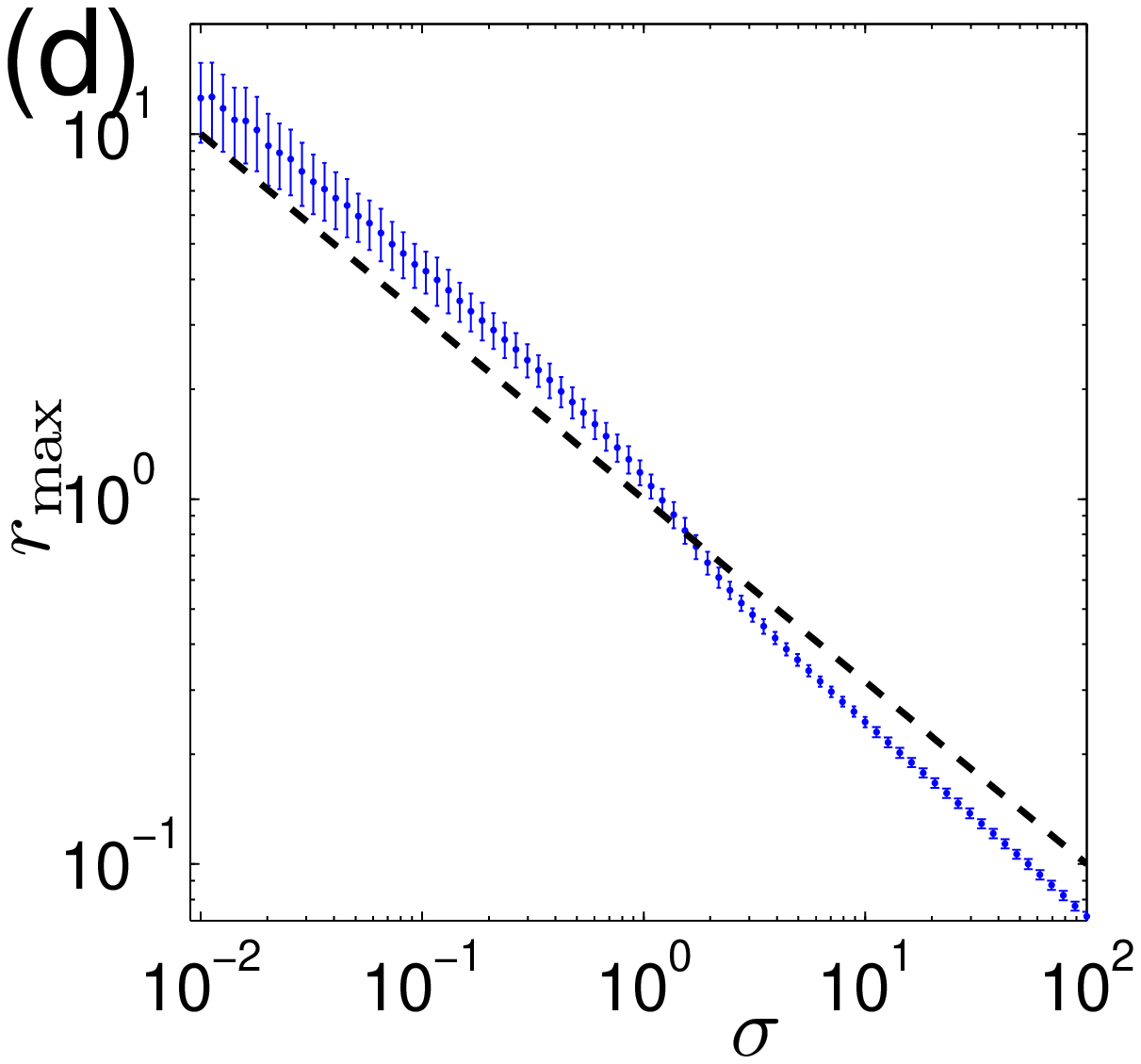} 
\caption{\small (Color online) (a) Unidirectional ring network with $L=30$ states and $E_{sc}=15$ random shortcuts. (b) Rearranging the nodes on the unit circle according to the argument of the complex eigenvector to eigenvalue $\lambda_{max}=-0.52 - 0.77\textnormal{i}$ of maximal ratio $r_{max}=1.48$ between imaginary and real part strong circular flows in the network can be detected. Forward transitions are shown in blue (solid arrows) and backward transitions in red (dashed arrows). (c) Convex hull of the complex eigenvalues excluding $\lambda_0=0$ for single realizations of directed ring networks of size $L=1000$ at various shortcut densities $\sigma=E_\textnormal{sc}/L$. Here $E_\textnormal{sc}$ is the number of additional random transition channels (including duplicates). The dashed lines originating at zero are tangents of slope $|r|=1.09$ to the convex hull for $\sigma=1.0$. The dots indicate nonzero complex eigenvalues for a network realization at $\sigma=2.0$. The spectral gap increases with $\sigma$ and the slope of the tangent decreases. (d) Double logarithmic plot of the average maximum value of $r=\textnormal{Im}[\lambda]/\textnormal{Re}[\lambda]$ in the spectrum versus shortcut density $\sigma$. Each data point and standard deviation was determined from a sample of $1000$ random networks of size $L=500$. The dashed line in (c) is $r=1/\sqrt{\sigma}$.}
\label{Fig:NetworkFlow}
\end{figure}
While the jump process on a ring is a discrete approximation for a limit-cycle oscillator with finite phase diffusion, this one-dimensional geometry may not reflect the topology of a more complicated oscillatory or transport process. A way to incorporate a more complex topology is to use a network of states that are not visited in sequence. Here we study the effect of adding $E_{sc}$ random shortcuts to a unidirectional ring of size $L$ (Fig.\ref{Fig:NetworkFlow}a). The resulting Markov chain is strongly connected ensuring that eigenvalue $\lambda_0=0$ is non-degenerate. 
\\ \\
The strength of the resonance depends on coherence ratio $r_k=\textnormal{Im}[\lambda_k]/\textnormal{Re}[\lambda_k]$ of resonant eigenmode $k$. The corresponding eigenvector $\mathbf{u}^{(k)}$ and its complex conjugate quantify collective oscillations and are associated with nonzero stationary probability fluxes in the system. Figure \ref{Fig:NetworkFlow}a shows a typical network for $L=30$, $E_\textnormal{sc}=15$, i.e., shortcut density $\sigma=0.5$, where each node $n$ is located on a circle at phase $\vartheta_n=2\pi n/L$. The same network is shown in Fig.\ref{Fig:NetworkFlow}b with $\vartheta_n = \textnormal{arg}(u^{(k)}_n)$, where $\mathbf{u}^{(k)}$ is the right eigenvector of $\textnormal{W}^0$ for the eigenvalue with the maximal ratio $r_k$. Forward transitions are highlighted blue (solid) and backward transitions red (thin, dashed). With respect to this arrangement, the flow is strongly biased in the forward direction.
\\ \\
We investigated the dependence of the expected maximum value $r_{\max}$ of the coherence ratio on shortcut density $\sigma=E_{sc}/L$. 
The eigenvalues of the transition matrices are localized in a region of the complex plane with a negative real part and are symmetric with respect to the line of real numbers (Fig.\ref{Fig:NetworkFlow}c).
The slope of a line originating at zero and tangent to that region scales with some power of $\sigma$ in both limits $\sigma\to 0$ and $\sigma\to\infty$ (Fig.\ref{Fig:NetworkFlow}d). 
The decrease in $r_{\max}$ with increasing $\sigma$ is the result of stronger mixing because of the random shortcuts. In the region of the topological crossover at $\sigma=1$ \cite{ToMaKo10} we observe that the coherence ratio is approximately equal to unity (see Figs.\ref{Fig:NetworkFlow}c and \ref{Fig:NetworkFlow}d). 
\begin{figure}[!t] 
%
 \includegraphics[width=4.25cm]{\picpath/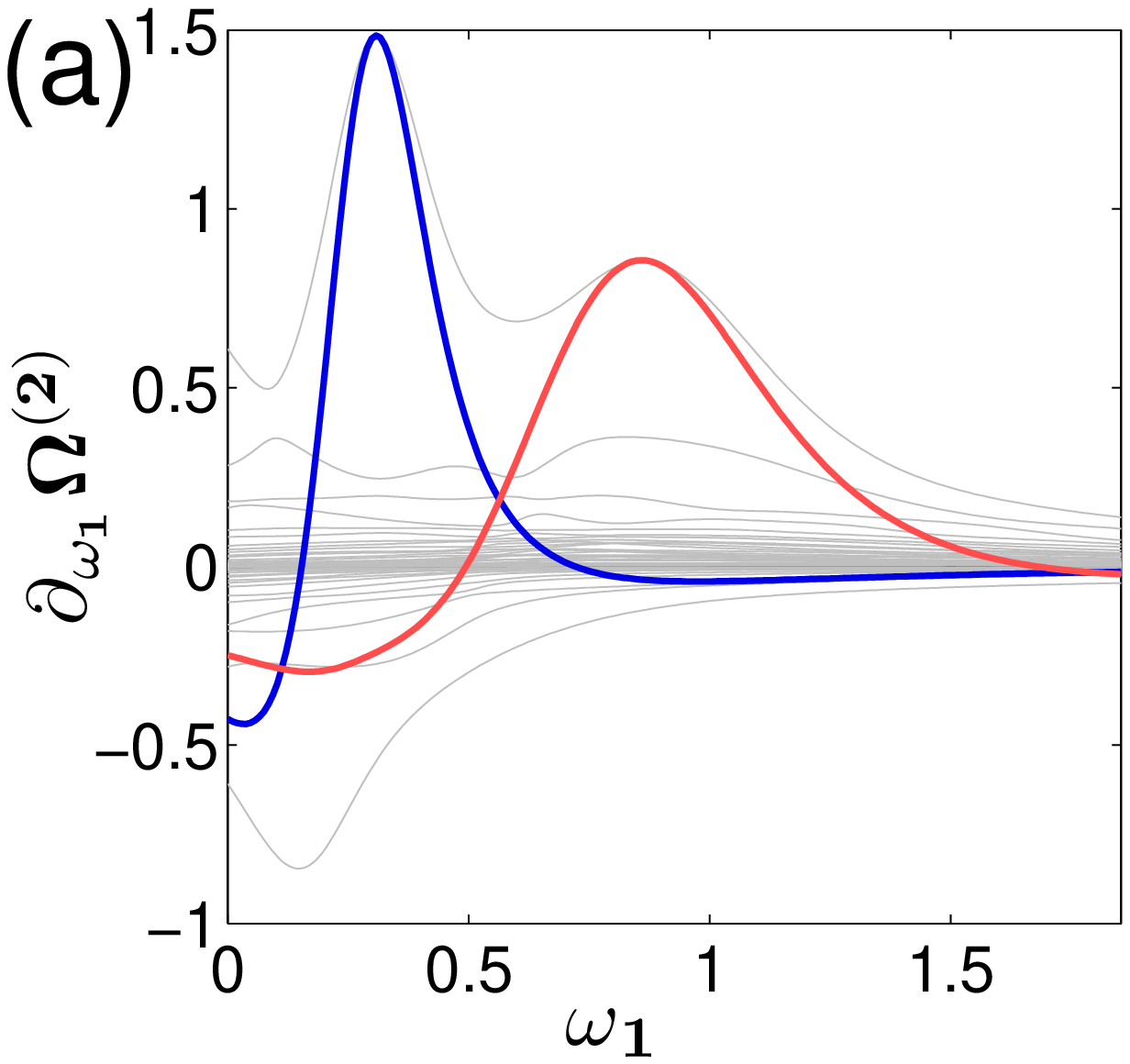}
 \includegraphics[width=4.25cm]{\picpath/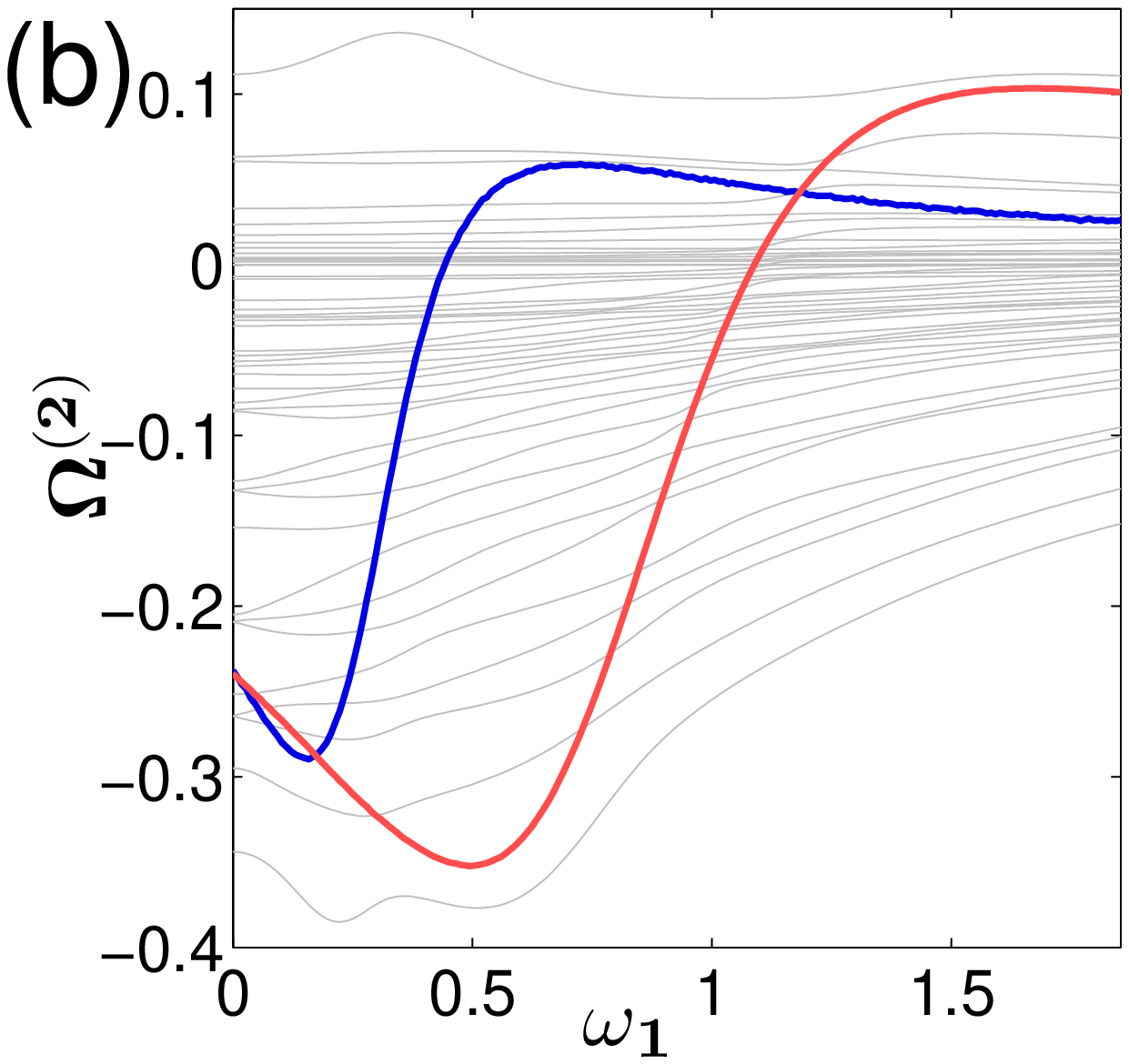} 
\caption{\small (Color online) Optimization of the driving protocol for the network shown in Fig.\ref{Fig:NetworkFlow}b. Assigning the same rate and perturbation cost to each transition in (a), we plot the eigenvalues of the derivative of the frequency shift operator (thin lines) with respect to frequency $\omega_1$ of deterministic harmonic driving. In (b), the eigenvalues of the second-order frequency-shift operator (thin lines) are shown as functions of the driving frequency. The red (bold, light gray) and blue (bold, dark gray) lines in (a) and (b) show the frequency shift and its derivative when the driving protocol is chosen to have the optimal frequency response at either of the two apparent resonant frequencies seen in (a).}
\label{Fig:NetworkResponse}
\end{figure}
\\ \\
When the transition rates are perturbed harmonically with frequency $\omega_1$ and driving protocol $\textnormal{H}$ that encodes the relative amplitudes and phases between the perturbations, the mean frequency will change by a value given approximately by Eqs.~(\ref{Eq:Hermdw01}) and (\ref{Eq:MainResult01}) in Sec.\ref{Sec:Pertheory}. Using the technique described in Sec.\ref{Sec:Opt} we optimized the driving protocol for the example shown in Fig.\ref{Fig:NetworkFlow}b such that, for a fixed driving frequency $\omega_1$, responsiveness $\frac{\partial\Omega^{(2)}}{\partial\omega_1}$ of the frequency shift is maximized. We assumed that the perturbation cost is the same for every transition.
We chose $\omega_1$ to be one of the apparent resonant frequencies in Fig.\ref{Fig:NetworkResponse}a. Synchronization, i.e., the adaptation of the mean frequency to the driving frequency, both positive and negative, is observed in Fig.\ref{Fig:NetworkResponse}b when these driving protocols are kept constant and $\omega_1$ is changed.
\section{Discussion}	\label{Sec:Outlook}
We have presented second-order perturbation analysis for the frequency of periodically or stochastically driven, time-continuous Markov chain models. In the case of a biased jump process on a ring, we also determined the second-order perturbation terms for the phase diffusion constant. This simplest discrete-state model of a stochastic oscillator can adapt its mean frequency toward the frequency of a deterministic or stochastic driving signal. The mechanism for this weak form of synchronization is a parametric resonance of the driving signal with an oscillatory relaxation mode of the stochastic process. We also showed that, when two identical stochastic oscillators are attractively coupled, they can increase their P\'eclet number, a measure for the coherence of oscillations, above the level of the uncoupled system. Furthermore, the explicit expressions for the frequency shift could be used to optimize the perturbations in a complex transition network.
\\ \\
Phase response can be viewed as a perturbation theory for the mean frequency of an oscillator. The techniques presented in this paper could therefore be useful in developing effective phase models for noise-induced or strongly noise-perturbed oscillations \cite{SchwaPiko10}. The explicit perturbation expressions for the frequency shift of a periodically driven Markov rate process may be used to estimate the unperturbed transition rates. In the case of enzymatic reactions, it may thus be possible to examine the reaction kinetics by measuring the response of the turnover rate to changes in the frequency and wave form of the driving signal.
\\ \\
Of course, the theory presented in this paper is also applicable to other nonequilibrium systems that are reasonably well described by stochastic cycles through the states of a Markov chain as well as stochastic transport or growth processes. In \cite{JOhkubo08}, periodic driving of the transition rates in the asymmetric exclusion process, which models transport through a cell membrane, was studied perturbatively. The resonance of the mean flow with the driving frequency in the second order in the perturbation strength was found, which agrees with our general results.
\\ \\
\\ \\
The authors thank Lutz Schimansky-Geier and Jun Ohkubo for helpful comments.
\section*{Appendix A}
In this section, we will derive general expressions for the time-dependent transition probabilities of a driven Markov rate process up to the second order $\varepsilon^2$ in perturbation strength. Transition probabilities $P_{nn_0}(t,t_0)$ obey the following equation
\beq	\label{EqA:MarkovChain}
	 \dot{ \textnormal{P}}(t|t_0) = \left[\textnormal{W}^0 + \varepsilon \textnormal{V}(t)\right] \textnormal{P}(t|t_0)
\eeq
with
\beq	\label{EqA:ComplexV}
	\textnormal{V}(t) = \textnormal{H} z(t) + \textnormal{H}^* z^*(t)	.
\eeq
We assume that complex driving signal $z(t)$ is stationary and symmetric such that
\besplit	\label{EqA:zCondition}
	\left\langle z(t+\tau)z(t)\right\rangle_t = \left\langle z(t)  \right\rangle_t = 0, \quad &\textnormal{and} \\
	c(\tau) = \left\langle z^*(t+\tau)z(t)\right\rangle_t = e^{\Lambda \tau}	\quad &\textnormal{for } \tau>0.					
\eesplit
The perturbation expansion of $\textnormal{P}(t|t_0)$ will be given in terms of eigenvectors and eigenvalues of unperturbed transition rates $\textnormal{W}^0$. We denote the left and right eigenvectors as $\mathbf{v}^{(k)}$ and $\mathbf{u}^{(k)}$, respectively, and the corresponding eigenvalues as $\lambda_k$. The completeness of and orthogonality relations for the eigenvectors are $\sum_{k} \mathbf{u}^{(k)}\mathbf{v}^{(k)\dagger}=\mathbb{1}$ and $\mathbf{v}^{(k)\dagger}\mathbf{u}^{(k')}=\delta_{kk'}$, respectively. In the main text we also set $v^{(0)}_n=1$ for $1\le n\le L$. If $\textnormal{W}^0$ is an infinite periodic difference operator, the eigenvalues are distributed continuously in a finite number of bands. Then, the completeness relation is expressed as a sum over different bands and for each band an integral over wave numbers $(-\pi,\pi]$. The orthogonality relation is given  by the product of a Kronecker delta for the band number and a Dirac delta function for the wave numbers within the bands. The nondegenerate zero eigenvalue corresponding to the unique stationary solution of the unperturbed problem is denoted as $\lambda_0=0$. By inserting the perturbation ansatz
\beq	\label{EqA:PertAnsatz}
    \textnormal{P}(t|t_0) = \textnormal{P}^{(0)}(t|t_0) + \sum_l \varepsilon^l \textnormal{P}^{(l)}(t|t_0)
\eeq
in Eq.~(\ref{EqA:MarkovChain}) and sorting by powers of $\varepsilon$, the dynamics of each perturbation term is given by a linear, inhomogeneous, ordinary differential equation
\beq	\label{EqA:PlODE}
	 \dot{\textnormal{P}}^{(l)}(t|t_0) = \textnormal{W}^0 \textnormal{P}^{(l)}(t|t_0) + \textnormal{V}(t) \textnormal{P}^{(l-1)}(t|t_0)	.
\eeq
In terms of Fourier modes
\besplit	\label{EqA:FourierModes}
      \pi^{(l)}_{kk'}(t|t_0) &= \mathbf{v}^{(k)\dagger} \textnormal{P}^{(l)}(t|t_0) \mathbf{u}^{(k')},			\\ 
      q^+_{kk'} 	&= \mathbf{v}^{(k)\dagger} \textnormal{H} \mathbf{u}^{(k')},			\\ 
      q^-_{kk'} 	&= \mathbf{v}^{(k)\dagger} \textnormal{H}^* \mathbf{u}^{(k')}			
\eesplit
and using the completeness relation, Eq.~(\ref{EqA:PlODE}) becomes
\beq	\label{EqA:FourierPlODE}
      {\dot \pi}^{(l)}_{kk'}	= \lambda_k \pi^{(l)}_{kk'}	+	\sum_{k''} \left[q^+_{kk''} z(t) + q^-_{kk''} z^*(t) \right] \pi^{(l-1)}_{k''k'}.
\eeq
Note that because $\mathbf{v}^{(0)\dagger}\textnormal{H} = \mathbf{v}^{(0)\dagger}\textnormal{H}^* = 0$, coefficients $q^\pm_{kk'}$ vanish for $k=0$. We will explicitly exempt these modes from the sum or integral in Eq.~(\ref{EqA:FourierPlODE}). From the initial condition $\textnormal{P}_{nm}(t|t) = \textnormal{P}_{nm}^{(0)}(t|t) = \delta_{nm}$ follows $\pi^{(l)}_{kk'}(t|t)=\delta_{l0}\delta_{kk'}$ and therefore
\beq	\label{EqA:SolZero}
      \pi_{kk'}^{(0)}(t + \tau|t) = e^{\lambda_k\tau} \delta_{kk'}. 
\eeq
All higher order perturbation terms are given by convolutions of the inhomogeneous part with an exponential kernel. Defining
\beq	\label{EqA:LaplaDef}
	\Laplace^t_{kk'}[f](\tau) = \int_0^\tau f(t+t') e^{-(\lambda_k-\lambda_{k'})t'}dt',
\eeq
the general solution to Eq.~(\ref{EqA:FourierPlODE}) for $l>0$ is
\besplit		\label{EqA:SolAll}
	\pi_{kk'}^{(l)}(t+\tau|t) &= e^{\lambda_k\tau} \sum_{k''} q^+_{kk''} \Laplace^t_{k0}\left[z\pi^{(l-1)}_{k''k'}\right](\tau)	\\ 
	&+ e^{\lambda_k\tau} \sum_{k''} q^-_{kk''} \Laplace^t_{k0}\left[z^*\pi^{(l-1)}_{k''k'}\right](\tau).	
\eesplit
The first- and second-order perturbation terms are thus, respectively,
\besplit	\label{EqA:SolOne}
	\pi_{kk'}^{(1)}(t+\tau|t) 	&= e^{\lambda_k \tau} q^+_{kk'} \Laplace^t_{kk'}[z](\tau) \\
					&+ e^{\lambda_k \tau} q^-_{kk'} \Laplace^t_{kk'}[z^*](\tau), 
\eesplit
and
\besplit \label{EqA:SolTwo}
	\pi_{kk'}^{(2)}(t+\tau|t) 	&= e^{\lambda_k \tau} \sum_{k''\ne 0} q^+_{kk''}q^-_{k''k'} \Laplace^t_{kk''}\left[z \Laplace^t_{k''k'}[z^*]\right](\tau) \\ 
					&+ e^{\lambda_k \tau} \sum_{k''\ne 0} q^-_{kk''}q^+_{k''k'} \Laplace^t_{kk''}\left[z^* \Laplace^t_{k''k'}[z]\right](\tau) \\ &+ O(z^2)	.
	\raisetag{20pt}
\eesplit
Here $O(z^2)$ denotes the products of the form $z(t+t')z(t+t'')$ and $z^*(t+t')z^*(t+t'')$ which will vanish when averaged over $t$.
The remaining nonvanishing time averages are expressed by incomplete Laplace transformations of complex autocorrelation function $c(\tau)=\left\langle z^*(t+\tau)z(t)\right\rangle_t$ or its complex conjugate. We observe

\besplit \label{EqA:zLzAvrg}
	&\left\langle z^*(t+\tau) \Laplace^t_{kk'}[z](\tau) \right\rangle_t  	\\ 
	&{\int_0^\tau \left\langle z^*(t+\tau) z(t+t') \right\rangle_t e^{-(\lambda_{k}-\lambda_{k'})t'}dt'}  \\  
	&= \int_0^\tau c(\tau-t') e^{-(\lambda_{k}-\lambda_{k'})t'}dt' \\
	& = \int_0^\tau c(t') e^{(\lambda_{k'}-\lambda_k)(\tau-t')}dt'   \\  
	&= e^{(\lambda_{k'}-\lambda_k)\tau}\Laplace_{k'k}^0[c](\tau)	
\eesplit
and

\besplit \label{EqA:LzLzAvrg}
	&\left\langle\Laplace^t_{kk''}\left[z^* \Laplace^t_{k''k'}[z]\right](\tau)\right\rangle_t  \\ 
	&=\int_0^\tau e^{(\lambda_{k'}-\lambda_{k''})t'} \Laplace_{k'k''}^0[c](t') ~e^{-(\lambda_{k}-\lambda_{k''})t'} dt'	 \\  
	&= \Laplace^0_{kk'}\left[\Laplace^0_{k'k''}[c]\right](\tau).		
\eesplit
For $c(\tau) = \exp(\Lambda \tau)$, the incomplete Laplace transformations are solved explicitly. For short hand notation, we define
\beq
\label{EqA:gkkL}
	g_{kk'}^\Lambda = \lambda_{k'}-\lambda_{k}+\Lambda.
\eeq
Then,
\beq
\label{EqA:ExplicitLaplace01}
      e^{\lambda_{k'}\tau}\Laplace_{k'k}^0[c](\tau) = \frac{e^{\lambda_{k'}\tau}}{g_{k'k}^\Lambda} \left(e^{g^\Lambda_{k'k}\tau}-1\right)
\eeq
and
\besplit   \label{EqA:ExplicitLaplace02}
      &e^{\lambda_{k}\tau}\Laplace^0_{kk'}\left[\Laplace^0_{k'k''}[c]\right](\tau)	\\
      &=e^{\lambda_k\tau} \Laplace^0_{kk'}\left[\frac{1}{g^\Lambda_{k'k''}}\left(e^{g^\Lambda_{k'k''}t}-1\right) \right] 	\\
      &= \frac{e^{\lambda_k\tau}}{g^\Lambda_{k'k''}}  \left(\frac{e^{g^\Lambda_{kk''}\tau}-1}{g^\Lambda_{kk''}}-\frac{e^{g^0_{kk'}\tau}-1}{\lambda_{k'}-\lambda_k}\right).
\eesplit

For $k'=k$, this equation takes the form
\beq	\label{EqA:ExplicitLaplace03}
      e^{\lambda_{k}\tau}\Laplace^0_{kk}\left[\Laplace^0_{kk''}[c]\right](\tau) = 
      \frac{e^{\lambda_k\tau}}{g^\Lambda_{kk''}} \left(\frac{e^{g^\Lambda_{kk''}\tau}-1}{g^\Lambda_{kk''}}-\tau\right).	
\eeq
For $k\ne 0$ and $k''\ne 0$, the dependence on the initial conditions is removed by taking limit $\tau\to\infty$ in Eqs.~(\ref{EqA:ExplicitLaplace01}) and (\ref{EqA:ExplicitLaplace02}):
\besplit
	\label{EqA:AsympLaplace}
      \lim_{\tau\to\infty} e^{\lambda_{k'}\tau}\Laplace_{k'k}^0[c](\tau) &= - \delta_{k'0} \frac{1}{g_{0k}^\Lambda},						\\
      \lim_{\tau\to\infty} e^{\lambda_{k}\tau}\Laplace^0_{kk'}\left[\Laplace^0_{k'k''}[c]\right](\tau) &= \delta_{k'0} \frac{1}{\lambda_k g^\Lambda_{0k''}}.	
\eesplit
The limits exist because the real parts of $\lambda_k$ and $\lambda_{k''}$ are strictly negative and the real part of $\Lambda$ is assumed to be less than or equal to zero.
Terms $\pi^{(l)}_{kk'}$ with $k'\ne 0$ vanish in the limit $\tau\to\infty$.
The explicit time dependence of the transition probabilities is necessary to obtain the generating function for random process $n(t)$. From the generating function, we can calculate both mean frequency and the asymptotic phase diffusion constant, as shown in Appendix B for the biased jump process on a ring. In the general case, the mean frequency can also be calculated from the asymptotic mean probability flows in the system. Let $\Theta_{nm}\in\lbrace -1,0,1\rbrace$ define for each transition whether it crosses a Poincar\'e section and in which direction. Mean frequency $\omega$ is proportional to the time averaged flow over this Poincar\'e section times $2\pi$,
\beq	\label{EqA:MeanOmegaGeneral}
	\Omega = 2\pi \sum_{n,m} \Theta_{nm} \left\langle J_{nm}(t) \right\rangle_t
\eeq
and
\besplit \label{EqA:MeanFlowGeneral}
	J_{nm}(t)		&=	W^0_{nm}p_m^{(0)} + 	\varepsilon \left(W^0_{nm}p_m^{(1)}(t) + V_{nm}(t)p_m^{(0)}\right)  		\\
				&+ 	\varepsilon^2 \left(W^0_{nm}p_m^{(2)}(t) + V_{nm}(t)p_m^{(1)}(t)\right)			. 
\eesplit
The time averages in the linear order of $\varepsilon$ are zero because of Eq.~(\ref{EqA:zCondition}).
We are now able to express the mean frequency in the second order of $\varepsilon^2$ by using Eqs.~(\ref{EqA:SolOne}),(\ref{EqA:SolTwo}),(\ref{EqA:ExplicitLaplace01}),(\ref{EqA:ExplicitLaplace02}), and (\ref{EqA:AsympLaplace}). We have
\besplit \label{EqA:Vp1}
	&\left\langle V_{nm}(t)p^{(1)}_m(t) \right\rangle_t 				\\
	&=\sum_{k \ne 0} u^{(k)}_m \left (H_{nm}  \left\langle z \pi^{(1)}_{k0}\right\rangle_t  + H^*_{nm} \left\langle z^* \pi^{(1)}_{k0}\right\rangle_t \right) v^{(0)*}_0 			\\
	&= - \sum_{k\ne 0} u^{(k)}_m \left( H_{nm} \frac{q^-_{k0}} {g^{\Lambda^*}_{0k}} 	+H_{nm}^*\frac{q^+_{k0}} {g^{\Lambda}_{0k}} \right) v^{(0)*}_0 	
	\raisetag{22pt}
\eesplit
as well as
\besplit \label{EqA:W0p2}
	&W^0_{nm} \left\langle p^{(2)}_m(t) \right\rangle_t = \sum_{k \ne 0} u^{(k)}_m W^0_{nm}  \left\langle \pi^{(2)}_{k0}\right\rangle_t v^{(0)*}_0 			\\
	&= \sum_{kk' \ne 0} u^{(k)}_m W^0_{nm} \left( \frac{q^+_{kk'}q^-_{k'0}}{\lambda_kg^{\Lambda^*}_{0k'}} + \frac{q^-_{kk'}q^+_{k'0}}{\lambda_kg^{\Lambda}_{0k'}}\right)v^{(0)*}_0.
\eesplit
Shift $\Omega^{(2)}$ in the mean frequency to the order $\varepsilon^2$ as a function of the complex driving protocol $\textnormal{H}$ is given in terms of the bilinear form $\nu(\textnormal{A}^*,\textnormal{B})$
\beq	\label{EqA:Hermdw}
	\Omega^{(2)}(\textnormal{H}^*,\textnormal{H}) = 2\pi \left[\nu(\textnormal{H}^*,\textnormal{H}) + \nu^*(\textnormal{H}^*,\textnormal{H})\right].
\eeq
The bilinear form $\nu(\textnormal{A}^*,\textnormal{B})$ is obtained from Eqs.~(\ref{EqA:gkkL}) and (\ref{EqA:MeanOmegaGeneral})-(\ref{EqA:W0p2}) as
\besplit
\label{EqA:nu2Form}
	  \nu (\textnormal{A}^*,\textnormal{B}) &= - \sum_{nm} \Theta_{nm} \sum_{k\ne 0} u^{(k)}_m v^{(0)}_0 \left( 
	  \vphantom{\sum_{k'\ne 0}}
	  A^*_{nm} \frac{\mathbf{v}^{(k)\dagger}\textnormal{B}\mathbf{u}^{(0)}}{\lambda_k + \Lambda}  	  \right. \\
	  &\left. - W^0_{nm} \sum_{k'\ne 0} \frac{\mathbf{v}^{(k)\dagger}\textnormal{A}^*\mathbf{u}^{(k')}\mathbf{v}^{(k')\dagger}\textnormal{B}\mathbf{u}^{(0)}}{\left(\lambda_{k'}+\Lambda\right)\lambda_k}  \right).
	  \raisetag{25pt}
\eesplit
If exponent $\Lambda$ depends in some way on frequency $\omega_1$ of complex driving signal $z(t)$, the responsiveness of the frequency shift to $\omega_1$ is calculated from
\besplit
\label{EqA:dnudW}
	  \partial_{\omega_1} \nu (\textnormal{A}^*,\textnormal{B}) &=  \partial_{\omega_1} \Lambda \sum_{nm} \Theta_{nm} \sum_{k\ne 0} u^{(k)}_m v^{(0)}_0
	  \left( \vphantom{\sum_{k'\ne 0}} A^*_{nm} \frac{\mathbf{v}^{(k)\dagger}\textnormal{B}\mathbf{u}^{(0)}}{(\lambda_k + \Lambda)^2}  	  \right. \\
	  &\left. - W^0_{nm} \sum_{k'\ne 0} \frac{\mathbf{v}^{(k)\dagger}\textnormal{A}^*\mathbf{u}^{(k')}\mathbf{v}^{(k')\dagger}\textnormal{B}\mathbf{u}^{(0)}}{\left(\lambda_{k'}+\Lambda\right)^2\lambda_k}  \right).
	  \raisetag{25pt}
\eesplit
\section*{Appendix B}
We consider a biased jump process on a ring of states with forward transition rates $W^0_{n+1,n} = L(1+\gamma)$ and backward transition rates $W^0_{n,n+1} = L\gamma$. The stationary probability flow in this system is $J^0_{n+1,n} = (W^0_{n+1,n} - W^0_{n,n+1})/L = 1$, i.e., the time scale is chosen such that the mean period is one and the mean frequency is $\omega_0=2\pi$. We split the transition rates into an unbiased jump process with forward and backward rates $W^{\textnormal{df}}_{n\pm 1,n}=L\gamma$ and bias $W^{\textnormal{fl}}_{n+1,n}=L$. The perturbation shall be given as a traveling wave to the forward transitions
\beq	\label{EqB:TravelWaveV}
	V_{nm}(t) = W^{\textnormal{fl}}_{nm} A(t)\cos(k_0 m - \phi(t)).
\eeq
Amplitude $A(t)$ and phase $\phi(t)$ of the perturbation specify complex driving signal $z(t)=A(t)\exp(-i\phi(t))$, which is supposed to be symmetric and posses a discrete or continuous rotational symmetry. Then, the eigenvectors of the unperturbed system are simple harmonics and the perturbation couples a mode $k$ to its neighboring modes $k\pm k_0$. The mean frequency in the second order in perturbation strength can be calculated from the expressions derived in Appendix A. Here we will use a different approach and derive the frequency shift as well as the asymptotic phase diffusion constant from the generating function of the biased jump process on an infinite lattice. In this case, eigenmodes
\beq	\label{EqB:Eigenvectors}
	v^{(k)}_n = 2\pi u^{(k)}_n = e^{ikn},	\quad k\in (-\pi,\pi]
\eeq
form a complete and orthogonal set with
\besplit \label{EqB:CompleteOttoNormal}
	&\int_{-\pi}^\pi v^{(k)*}_m u_n^{(k)}  dk = \delta_{nm}	\qquad \textnormal{and}  \\ \\
	&\mathbf{v}^{(k)\dagger}\mathbf{u}^{(k')} = \sum_{n=-\infty}^\infty v^{(k)*}_n u_n^{(k')} = \delta(k-k')	.		
\eesplit
The eigenvalues follow from
\besplit \label{EqB:EigenEquation}
	&\sum_{m=-\infty}^\infty W^0_{nm} u^{(k)}_m = \lambda_k u^{(k)}_n 	\\
	&= L(1+\gamma) \left(u^{(k)}_{n-1} - u^{(k)}_n\right) + L\gamma \left(u^{(k)}_{n+1} - u^{(k)}_n\right) \\
	&= \left[ L(1+\gamma) \left(e^{-ik} - 1\right) + L\gamma \left(e^{ik} - 1\right)\right] u^{(k)}_n. 
\eesplit
They can be divided into diffusion and flow parts as
\besplit \label{EqB:LambdaDecomp}
      \lambda_{k} &= \lambda^{\textnormal{fl}}_k + \lambda^{\textnormal{df}}_k, 	\\
      \lambda^{\textnormal{fl}}_k &= L\left(e^{-ik}-1\right),			\\
      \lambda^{\textnormal{df}}_k &= L\gamma \left(e^{ik}+e^{-ik}-2\right).
\eesplit
Perturbation $\textnormal{V} = \textnormal{H}z(t) + \textnormal{H}^*z^*(t)$ applied to mode $\mathbf{u}^{(k)}$ creates two neighboring modes $\mathbf{u}^{(k\pm k_0)}$ as
\besplit	\label{EqB:PerturbAction}
      \textnormal{H} \mathbf{u}^{(k)} &= \frac{1}{2}\lambda^{\textnormal{fl}}_{k+k_0} \mathbf{u}^{(k+k_0)}, \\
      \textnormal{H}^* \mathbf{u}^{(k)} &= \frac{1}{2} \lambda^{\textnormal{fl}}_{k-k_0} \mathbf{u}^{(k-k_0)},
\eesplit
so that
\besplit	\label{EqB:FourierModes}
     q^+_{kk'} &= \mathbf{v}^{(k)\dagger}\textnormal{H}\mathbf{u}^{(k')} = \frac{1}{2}\lambda^{\textnormal{fl}}_{k} \delta(k-k'-k_0), \\
     q^-_{kk'} &= \mathbf{v}^{(k)\dagger}\textnormal{H}^*\mathbf{u}^{(k')} = \frac{1}{2}\lambda^{\textnormal{fl}}_{k} \delta(k-k'+k_0).
\eesplit
The mean frequency and the phase diffusion constant are defined as
\besplit
\label{EqB:MeanMomentLimits01}
	\Omega &= \frac{2\pi}{L} \lim_{\tau\to\infty} \frac{1}{\tau}\mathbb{E}[n(t+\tau)-n(t)],	\\
	D &= \left(\frac{2\pi}{L}\right)^2 \lim_{\tau\to\infty} \frac{1}{2\tau}\textnormal{Var}([n(t+\tau)-n(t)).
\eesplit
Taking the long time limit makes $\omega$ and $D$ independent from the initial conditions. Therefore, we can start with a localized distribution at $n(t)=0$ and average over time, i.e., study at the asymptotic behavior of the moments of the time-averaged transition probability $\left\langle P_{n0}(t+\tau|t) \right\rangle_t$. These moments are conveniently calculated from the characteristic function
\beq	\label{EqB:MGF01}
	\phi_{\tau}(x) = \mathbb{E}\left[e^{ixn}\right] = \sum_{n} e^{ixn} \left\langle P_{n0}(t+\tau|t)\right\rangle_t
\eeq
as
\beq	\label{EqB:MeanMoments}
	\mathbb{E}[n] = -i\phi'(0),	\quad\textnormal{and} \quad \textnormal{Var}[n] = -\phi''(0) + \phi'^2(0).
\eeq
The perturbation expansion of the time-averaged transition probabilities is
\beq	\label{EqB:MeanP}
	\left\langle P_{n0}(t+\tau|t)\right\rangle_t = P_{n0}^{(0)}(\tau) + \varepsilon^2 \left\langle P^{(2)}_{n0}(t+\tau|t)\right\rangle_t.
\eeq
In Appendix A, we have already derived the Fourier modes of the transition probabilities up to the second order of $\varepsilon^2$.

\besplit \label{EqB:MGF02}
	  &\phi_{\tau}(x) = \sum_{n} e^{ixn} \left\langle P_{n0}(t+\tau|t)\right\rangle_t	\\ 
	  &= \int_{-\pi}^\pi \sum_n e^{ixn} u^{k}_n \left(\left\langle \pi^{(0)}_{kk'} \right\rangle_t + \varepsilon^2 \left\langle \pi^{(2)}_{kk'} \right\rangle_t \right) v^{(k')*}_0 dkdk'	\\
	  &= e^{\lambda_{-x}\tau} + \varepsilon^2 \int_{-\pi}^\pi \left\langle \pi^{(2)}_{-xk'} \right\rangle_t dk' .
	  \raisetag{80pt}
\eesplit
Here, we have used 
\beq
\label{EqB:DeltaDelta}
	\frac{1}{2\pi} \sum_{n=-\infty}^\infty e^{i(k+x)n} = \delta(k+x), \quad
	\pi^0_{kk'} = e^{\lambda_k} \delta(k-k').
\eeq
From Eq.~(\ref{EqA:SolTwo}), we see that the second-order perturbation term of average Fourier coefficient $\left\langle \pi^{(2)}_{-xk'} \right\rangle_t$ contains products of the form $q^+_{kk''}q^-_{k''k'}$, and with Eq.~(\ref{EqB:FourierModes}), it follows that these terms are nonzero only if $k=k'$ and \mbox{$k''=k\mp k_0$}. All integrals over the Fourier modes are thus evaluated explicitly, and we find
\besplit	\label{EqB:MGF2ndOrder}
	  &\int_{-\pi}^\pi \left\langle \pi^{(2)}_{-xk'} \right\rangle_t dk'  			\\
	  &=\frac{1}{4} e^{\lambda_{-x}\tau} \frac{\lambda^{\textnormal{fl}}_{-x+k_0}\lambda^{\textnormal{fl}}_{-x}}{g^{\Lambda}_{-x,-x+k_0}} \left(\frac{e^{g^{\Lambda}_{-x,-x+k_0}\tau}-1}{g^{\Lambda}_{-x,-x+k_0}}-\tau\right)	 \\
	  &+\frac{1}{4} e^{\lambda_{-x}\tau} \frac{\lambda^{\textnormal{fl}}_{-x-k_0}\lambda^{\textnormal{fl}}_{-x}}{g^{\Lambda^*}_{-x,-x-k_0}} \left(\frac{e^{g^{\Lambda^*}_{-x,-x-k_0}\tau}-1}{g^{\Lambda^*}_{-x,-x-k_0}}-\tau\right)	.
\eesplit
Inserting Eq.~(\ref{EqB:MGF2ndOrder}) into Eq.~(\ref{EqB:MGF02}), we find that the generating function has the form
\beq	\label{EqB:MGFStructure}
	\phi(x) = e^{\lambda_{-x}\tau} \left(1 + \varepsilon^2 h(x) \right).
\eeq
To emphasize the structure of $h(x)$ and take the derivative, we introduce some more shorthand notations
\besplit	\label{EqB:h1h2Def}
	h_1^+(x) &=\frac{1}{4}~\frac{\lambda^{\textnormal{fl}}_{-x+k_0}}{g^\Lambda_{-x,-x+k_0}}, \\ 
	h_2^+(x) &=\lambda^{\textnormal{fl}}_{-x} \frac{e^{g^\Lambda_{-x,-x+k_0}\tau}-1}{g^\Lambda_{-x,-x+k_0}}, \\
	h_1^-(x) &= \frac{1}{4}~\frac{\lambda^{\textnormal{fl}}_{-x-k_0}}{g^{\Lambda^*}_{-x,-x-k_0}}, \\
	h_2^-(x) &= \lambda^{\textnormal{fl}}_{-x} \frac{e^{g^{\Lambda^*}_{-x,-x-k_0}\tau}-1}{g^{\Lambda^*}_{-x~-x-k_0}}, 
\eesplit
so that
\besplit	\label{EqB:hStructure}
	h(x) =& h_1^+(x)h_2^+(x) + h_1^-(x)h_2^-(x) \\ &-(h_1^+(x) + h_1^-(x))  \lambda^{\textnormal{fl}}_{-x} \tau . 
\eesplit
The necessary derivatives of the eigenvalues are
\besplit	\label{EqB:MomentsDerive}
	\left.\frac{d}{dx}\lambda^{\textnormal{fl}}_{-x+k_0}\right|_{x=0} &= i \left(\lambda^{\textnormal{fl}}_{k_0} + L\right),	 \\
	\left.\frac{d^2}{dx^2}\lambda^{\textnormal{fl}}_{-x+k_0}\right|_{x=0} &= -\left(\lambda^{\textnormal{fl}}_{k_0} + L\right), \\
	\left.\frac{d}{dx}\lambda^{\textnormal{df}}_{-x+k_0}\right|_{x=0} &= -2\gamma \textnormal{Im}\left[\lambda^{\textnormal{fl}}_{k_0}\right], \\
	\left.\frac{d^2}{dx^2}\lambda^{\textnormal{df}}_{-x+k_0}\right|_{x=0} &= -\left(\lambda^{\textnormal{df}}_{k_0} + 2\gamma L \right),	
\eesplit
and with Eqs.~(\ref{EqB:MGFStructure}) and (\ref{EqB:MeanMoments}), it follows that
\besplit \label{EqB:MeanMomentLimits02}
	\mathbb{E}[n] &= L\tau - \varepsilon^2 ih'(0)	\qquad\textnormal{and}  \\ \\
	\textnormal{Var}[n] &= L(1+2\gamma)\tau - \varepsilon^2 h''(0).	
\eesplit
Functions $h^\pm_2(x)$ and all their derivatives at zero vanish when divided by $\tau$ in the limit $\tau\to\infty$
\beq	\label{EqB:h2ZeroLimits}
	\lim_{\tau\to\infty} \frac{1}{\tau} \left. \frac{d^n}{dx^n}h^\pm_2(x) \right|_{x=0}=0, \quad (n=0,1,\dots).
\eeq
Only derivatives of  $h^\pm_1(x)\lambda^{\textnormal{fl}}_{-x}\tau$ at zero remain in that limit. With Eq.~(\ref{EqB:h1h2Def}) and recalling that $g^\Lambda_{kk'}=\lambda_{k'}-\lambda_k + \Lambda$, we have
\besplit
\label{EqB:h1ZeroLimits}
	h_1^+(0) &=\frac{1}{4}~\frac{\lambda^{\textnormal{fl}}_{k_0}}{\lambda_{k_0}+\Lambda}, 	\\
	h_1^{+}{'}(0) & = i\frac{1}{4} \left[\frac{\lambda^{\textnormal{fl}}_{k_0}+L}{\lambda_{k_0}+\Lambda}-\frac{\lambda^{\textnormal{fl}}_{k_0}\left(\lambda^{\textnormal{fl}}_{k_0}+i 2\gamma \textnormal{Im}\left[\lambda^{\textnormal{fl}}_{k_0}\right]\right)}{(\lambda_{k_0}+\Lambda)^2}\right],
	\raisetag{50pt}
\eesplit
and $h_1^-(0) = h_1^{+}(0)^*$ and $h^-_1{'}(0)  = h^{+}_1{'}(0)^*$. Finally, we can collect all the terms necessary to write explicit expressions for the mean phase velocity, the phase diffusion constant, and their ratio the P\'eclet number
\besplit
\label{EqB:MeanOmegaRing}
	\Omega &= \frac{2\pi}{L} \lim_{\tau\to\infty} \frac{1}{\tau}\mathbb{E}[n] = 2\pi - \varepsilon^2 \frac{2\pi}{L}\lim_{\tau\to\infty} \frac{1}{\tau}ih'(0)	 \\ 
	&= 2\pi + \varepsilon^2 \frac{2\pi}{L} (h^+_1(0)+h^-_1(0))i\left.\frac{d}{dx}\lambda^{\textnormal{fl}}_{-x}\right|_{x=0}  \\  
	&= 2\pi \left( 1 - \varepsilon^2 \frac{1}{4}\left[\frac{\lambda^{\textnormal{fl}}_{k_0}}{\lambda_{k_0}+\Lambda} + \frac{\lambda^{\textnormal{fl}}_{-k_0}}{\lambda_{-k_0}+\Lambda^*}\right]\right)
\eesplit
and
\besplit
      D & = \left(\frac{2\pi}{L}\right)^2 \lim_{\tau\to\infty} \frac{1}{2\tau} \textnormal{Var}[n] \\
	& = \frac{2\pi^2}{L} (1+2\gamma) - \varepsilon^2 \left(\frac{2\pi}{L}\right)^2\lim_{\tau\to\infty} \frac{1}{2\tau}h''(0),
\eesplit
so that $D_0=(1+2\gamma) 2\pi^2/L$ and

\bmul \label{EqB:MeanPhaseDiffRing}
	\frac{D}{D_0} = 1 + \varepsilon^2 \frac{1}{L(1+2\gamma)} \left[ 
	\vphantom{2\frac{\lambda^{\textnormal{fl}}_{k_0}\left(\lambda^{\textnormal{fl}}_{k_0}+i2\gamma \textnormal{Im}\left[\lambda^{\textnormal{fl}}_{k_0}\right]\right)}{(\lambda_{k_0}+\Lambda)^2}}
									  2 h'^+_1(0) \left.\frac{d}{dx}\lambda^{\textnormal{fl}}_{-x}\right|_{x=0} \right. \\
								  \left.  + h^+_1(0)\left.\frac{d^2}{dx^2}\lambda^{\textnormal{fl}}_{-x}\right|_{x=0} +  c.c.
	\vphantom{2\frac{\lambda^{\textnormal{fl}}_{k_0}\left(\lambda^{\textnormal{fl}}_{k_0}+i2\gamma \textnormal{Im}\left[\lambda^{\textnormal{fl}}_{k_0}\right]\right)}{(\lambda_{k_0}+\Lambda)^2}}
								  \right]	\\
	\shoveleft = 1 + \varepsilon^2 \frac{1}{(1+2\gamma)} \left[ 
	\vphantom{2\frac{\lambda^{\textnormal{fl}}_{k_0}\left(\lambda^{\textnormal{fl}}_{k_0}+i2\gamma \textnormal{Im}\left[\lambda^{\textnormal{fl}}_{k_0}\right]\right)}{(\lambda_{k_0}+\Lambda)^2}}
								      2ih'^+_1(0) - h^+_1(0) +  c.c.
							     \right]	\\
	 \shoveleft = 1 + \varepsilon^2 \frac{1}{4(1+2\gamma)} \left[
							  2\frac{\lambda^{\textnormal{fl}}_{k_0}\left(\lambda^{\textnormal{fl}}_{k_0}+i2\gamma \textnormal{Im}\left[\lambda^{\textnormal{fl}}_{k_0}\right]\right)}{(\lambda_{k_0}+\Lambda)^2} \right. \\
						   \left. 
	\vphantom{2\frac{\lambda^{\textnormal{fl}}_{k_0}\left(\lambda^{\textnormal{fl}}_{k_0}+i2\gamma \textnormal{Im}\left[\lambda^{\textnormal{fl}}_{k_0}\right]\right)}{(\lambda_{k_0}+\Lambda)^2}}
						   - \frac{3\lambda^{\textnormal{fl}}_{k_0}+2L}{\lambda_{k_0}+\Lambda} + c.c. \right].
\emul
The relative P\'eclet number is 
\besplit
\label{EqB:MeanCoherenceRing}
	\frac{c}{c_0} &= 1 + \varepsilon^2 \left[\frac{\Omega^{(2)}}{\omega_0}-\frac{D^{(2)}}{D_0}\right] \\ 
	  & = 1 + \varepsilon^2 \frac{1}{2(1+2\gamma)}\left[\frac{(1-\gamma)\lambda^{\textnormal{fl}}_{k_0}+L}{\lambda_{k_0}+\Lambda} \right. \\ 
	  & \left.- \frac{\lambda^{\textnormal{fl}}_{k_0}\left(\lambda^{\textnormal{fl}}_{k_0}+i2\gamma\textnormal{Im}\left[\lambda^{\textnormal{fl}}_{k_0}\right]\right)}{(\lambda_{k_0}+\Lambda)^2} + c.c.\right],
\eesplit
where $c.c.$ denotes the complex conjugated terms.
\\ \\
From the generating function Eq.~(\ref{EqB:MGF01}) also follows the autocorrelation of the driving signal if it is taken from another stochastic oscillator as $z(t) = \exp(-ikn_1(t))$ :
\beq	\label{EqB:AutoCorr01}
	\left\langle z^*(t+\tau|t)z(t)\right\rangle_t = \sum_{n} e^{ikn} \left\langle P_{n0}(t+\tau|t) \right\rangle_t,
\eeq
and with Eqs.~(\ref{EqB:MGF01}) and (\ref{EqB:MGFStructure}), we find
\beq	\label{EqB:AutoCorr02}
	\left\langle z^*(t+\tau|t)z(t)\right\rangle_t = \phi(k) = e^{\lambda_{-k}\tau},
\eeq
i.e., $\Lambda=\lambda_{-k}=\lambda^*_{k}$ in Eq.~(\ref{EqA:zCondition}). Because the time scale of the driving signal can be different from the perturbed stochastic oscillator, eigenvalue $\lambda_{-k}$ is not necessarily given by Eq.~(\ref{EqB:LambdaDecomp}).
\\ \\
The continuum limit can be taken if $L,\gamma\to\infty$ and 
\beq	\label{EqB:ContLimD0}
	D_0 = \frac{2\pi^2}{L}(1+2\gamma)
\eeq
is kept constant. For harmonic driving, angle \mbox{$\vartheta=2\pi n/L$} in the continuum limit evolves according to
\beq	\label{EqB:ContLimDyn}
      \dot\vartheta = 2\pi (1 + \varepsilon \cos(\vartheta - \omega_1 t)) + \sqrt{2 D_0} \xi(t),
\eeq
where $\xi(t)$ is white noise with $\left\langle\xi(t)\xi(t')\right\rangle=\delta(t-t')$. After substituting $\varphi=\vartheta-\omega_1 t$, we obtain the stochastic Adler equation
\beq	\label{EqB:ContLimAdler}
      \dot\varphi = 2\pi-\omega_1 + 2\pi\varepsilon \cos(\varphi) + \sqrt{2 D_0} \xi(t)
\eeq
for which the mean frequency and the phase diffusion constant are known \cite{SchwaPiko10,VdBHaenggi01}. With $k_0=2\pi/L$ and $\Delta=\omega_1-2\pi$, the eigenvalues and the perturbation terms become
\bsubeq	
      \lambda^{\textnormal{fl}}_{k_0} &\to -i 2\pi, \qquad \lambda^{\textnormal{df}}_{k_0} \to - D_0,	\label{EqB:ContLimLam}	\\
      \frac{\Omega}{\omega_0} &\to 1 +  \varepsilon^2  \pi \frac{\Delta}{D_0^2+\Delta^2},	\label{EqB:ContLimOhm} \\
      \frac{D}{D_0} &\to 1 + \varepsilon^2 \left[ \frac{(2\pi)^2(\Delta^2-D_0^2)}{(D_0^2+\Delta^2)^2} + \frac{2\pi^2}{D_0^2 + \Delta^2} \right].	\label{EqB:ContLimD}
\esubeq
\section*{Appendix C}
Here we consider two stochastic jump processes $n_0(t)$ and $n_1(t)$ on a two-dimensional, periodic lattice. While we have used the time-dependent perturbation theory in the case of external driving, the coupled system $\mathbf{n}=(n_0,n_1)=n_0 \mathbf{e}^0 + n_1 \mathbf{e}^1$ is autonomous. Nevertheless, because the time-independent perturbation theory is a special case of the time dependent perturbation theory, we can set $z(t)=1$ and follow the derivations in Appendix B with a slight modification. The time averages have to be replaced with a suitable average over initial state $\mathbf{n^0}=\mathbf{n}(0)$. For this, we introduce unitary translation operator $\textnormal{T}_\mathbf{n^0}$ with $\left[\textnormal{T}^\dagger_{\mathbf{n^0}} \textnormal{P}\textnormal{T}_{\mathbf{n^0}}\right]_\mathbf{nm}=P_{\mathbf{n^0}+\mathbf{n},\mathbf{n^0}+\mathbf{m}}$. Transition probabilities $\textnormal{P}(\tau)=\textnormal{P}(t+\tau|t)$ are now independent of $t$, but Fourier modes
\beq
	\pi_{\mathbf{k}\mathbf{k'}}(\mathbf{n^0,\tau}) = \mathbf{v}^{(\mathbf{k})\dagger} \textnormal{T}^\dagger_{\mathbf{n^0}} \textnormal{P}(\tau)\textnormal{T}_{\mathbf{n^0}} \mathbf{u}^{(\mathbf{k'})}
\eeq
and
\besplit
	q^+_{\mathbf{kk'}}(\mathbf{n^0}) &= \mathbf{v}^{(\mathbf{k})\dagger} \textnormal{T}^\dagger_{\mathbf{n^0}} \textnormal{H}\textnormal{T}_{\mathbf{n^0}} \mathbf{u}^{(\mathbf{k'})}, \\
	q^-_{\mathbf{kk'}}(\mathbf{n^0}) &= \mathbf{v}^{(\mathbf{k})\dagger} \textnormal{T}^\dagger_{\mathbf{n^0}} \textnormal{H}^*\textnormal{T}_{\mathbf{n^0}} \mathbf{u}^{(\mathbf{k'})} 
\eesplit
depend explicitly on the initial conditions. Due to translational symmetry, transition matrix $\textnormal{W}^{0}$ and the translation operator $\textnormal{T}_\mathbf{n^0}$ commute.
\\ \\
Again, we divide the transition rates into isotropic diffusion part $\textnormal{W}^{\textnormal{df}}$ and a bias in the forward directions.
\besplit	\label{EqC:RateDecomp}
	\textnormal{W}^0 &= \textnormal{W}^{\textnormal{df}} + \textnormal{W}^{\textnormal{fl}},	\\
	W^{\textnormal{df}}_{\mathbf{n}\pm\mathbf{e^0},\mathbf{n}} &= W^{\textnormal{df}}_{\mathbf{n}\pm\mathbf{e^1},\mathbf{n}} = L\gamma,	\\
	W^{\textnormal{fl}}_{\mathbf{n}+\mathbf{e^0},\mathbf{n}} &= L	, \quad 	W^{\textnormal{fl}}_{\mathbf{n}+\mathbf{e^1},\mathbf{n}} = \frac{\omega_1}{2\pi} L.
\eesplit
The oscillators are coupled weakly through their forward jump rates with the perturbation
\besplit	\label{EqC:PertForm}
	V_{\mathbf{n}+\mathbf{e^0}, \mathbf{n}} &= L \cos(\mathbf{k^0}\mathbf{n}+\beta),	\\
	V_{\mathbf{n}+\mathbf{e^1}, \mathbf{n}} &= \frac{\omega_1}{2\pi} L \cos(\mathbf{k^0}\mathbf{n}-\beta).
\eesplit
Here $\mathbf{k^0}\mathbf{n}$ denotes the inner product. The eigenfunctions of the unperturbed system are harmonics \mbox{$v_\mathbf{n}^{(\mathbf{k})} = (2\pi)^2 u_\mathbf{n}^{(\mathbf{k})} = e^{i\mathbf{k}\mathbf{n}}$}. With $\mathbf{k}=(k_0,k_1)$, we define
\besplit	\label{EqC:LambdaDecomp}
	\lambda^{\textnormal{fl}}_k &= L \left(e^{-ik}-1\right), \\
	\lambda^{\textnormal{df}}_k &= L \left(e^{ik}+e^{-ik}-2\right), \\
	\lambda^{\textnormal{df}}_{\mathbf{k}} &= \lambda^{\textnormal{df}}_{k_0} + \lambda^{\textnormal{df}}_{k_1}, \\ 
	\lambda^{\textnormal{fl}}_{\mathbf{k}}(\beta) &= \lambda^{\textnormal{fl}}_{k_0} + \frac{\omega_1}{2\pi}\lambda^{\textnormal{fl}}_{k_1} e^{-i2\beta},	\\
	\lambda_{\mathbf{k}} &= \lambda^{\textnormal{fl}}_{\mathbf{k}}(0) + \lambda^{\textnormal{df}}_{\mathbf{k}}.
\eesplit
Then
\besplit	\label{EqC:PertAction}
	\textnormal{W}^0 \mathbf{u}^{(\mathbf{k})} &= \lambda_{\mathbf{k}} \mathbf{u}^{(\mathbf{k})},	\\
	\textnormal{H} \mathbf{u}^{(\mathbf{k})} & = \frac{1}{2}e^{i\beta}\lambda^{\textnormal{fl}}_{\mathbf{k}+\mathbf{k^0}}(\beta) \mathbf{u}^{(\mathbf{k}+\mathbf{k^0})}, \\
	\textnormal{H}^* \mathbf{u}^{(\mathbf{k})} & = \frac{1}{2}e^{-i\beta}\lambda^{\textnormal{fl}}_{\mathbf{k}-\mathbf{k^0}}(-\beta) \mathbf{u}^{(\mathbf{k}-\mathbf{k^0})}
\eesplit
and
\besplit	\label{EqC:PertFourier}
	q^+_{\mathbf{k}\mathbf{k'}} &= e^{i\mathbf{k^0n^0}} ~ \frac{1}{2} e^{i\beta} \lambda^{\textnormal{fl}}_{\mathbf{k}}(\beta) \delta(\mathbf{k}-\mathbf{k'}-\mathbf{k^0}),	\\
	q^-_{\mathbf{k}\mathbf{k'}} &= e^{-i\mathbf{k^0n^0}} ~ \frac{1}{2} e^{-i\beta} \lambda^{\textnormal{fl}}_{\mathbf{k}}(-\beta) \delta(\mathbf{k}-\mathbf{k'}+\mathbf{k^0}).	\\
\eesplit
We see that the average over the initial conditions in Eqs.~(\ref{EqA:SolOne}) and (\ref{EqA:SolTwo}) removes all terms linear in $q^\pm_{\mathbf{kk'}}$ and quadratic terms $q^+_{\mathbf{kk''}}q^+_{\mathbf{k''k'}}$ and $q^-_{\mathbf{kk''}}q^-_{\mathbf{k''k'}}$.
The generating function is written as
\beqarr	\label{EqC:MGF01}
	\phi(\mathbf{x}) 	&=& \sum_{\mathbf{n}} e^{i\mathbf{x}\mathbf{n}} \left\langle P_{\mathbf{n^0}+\mathbf{n},\mathbf{n^0}}(\tau)\right\rangle_\mathbf{n^0}	\nonumber \\
				&=& e^{\lambda_{-\mathbf{x}}\tau} + \varepsilon^2 \int_{-\pi}^\pi d\mathbf{k'} \left\langle \pi^{(2)}_{-\mathbf{x} \mathbf{k'}}\right\rangle_\mathbf{n^0}
\eeqarr
and from Eqs.~(\ref{EqA:SolTwo}),(\ref{EqA:ExplicitLaplace03}), and (\ref{EqC:PertFourier}) and setting $\Lambda=0$, we have
\besplit	\label{EqC:MGF2ndOrder}
	  &\int_{-\pi}^\pi \left\langle \pi^{(2)}_{-\mathbf{x} \mathbf{k'}} \right\rangle_\mathbf{n^0} d\mathbf{k'}  			\\
	  &=\frac{1}{4} e^{\lambda_{-\mathbf{x}}\tau} \frac{\lambda^{\textnormal{fl}}_{-\mathbf{x}+\mathbf{k^0}}(-\beta)\lambda^{\textnormal{fl}}_{-\mathbf{x}}(\beta)}{\lambda_{-\mathbf{x}+\mathbf{k^0}}-\lambda_{-\mathbf{x}}}\left(\frac{e^{(\lambda_{-\mathbf{x}+\mathbf{k^0}}-\lambda_{-\mathbf{x}})\tau}-1}{\lambda_{-\mathbf{x}+\mathbf{k^0}}-\lambda_{-\mathbf{x}}}-\tau\right)	 \\
	  &+\frac{1}{4} e^{\lambda_{-\mathbf{x}}\tau} \frac{\lambda^{\textnormal{fl}}_{-\mathbf{x}-\mathbf{k^0}}(\beta)\lambda^{\textnormal{fl}}_{-\mathbf{x}}(-\beta)}{\lambda_{-\mathbf{x}-\mathbf{k^0}}-\lambda_{-\mathbf{x}}} \left(\frac{e^{(\lambda_{-\mathbf{x}-\mathbf{k^0}}-\lambda_{-\mathbf{x}})\tau}-1}{\lambda_{-\mathbf{x}-\mathbf{k^0}}-\lambda_{-\mathbf{x}}}-\tau\right)	.
	  \raisetag{90pt}
\eesplit
As in the previous section, the generating function is of the form
\beq	\label{EqC:MGF02}
	\phi(\mathbf{x}) = e^{\lambda_{-\mathbf{x}}\tau} \left(1 + \varepsilon^2 h(\mathbf{x})\right)
\eeq
and the mean and variance of the position of the first oscillator are given, respectively, by
\besplit	\label{EqC:Moments}
	\mathbb{E}[n_0] &= -i\left.\partial_{x_0}\phi(\mathbf{x})\right|_{\mathbf{x}=\mathbf{0}},	\\
	\textnormal{Var}[n_0] &= -\left.\partial^2_{x_0} \phi(\mathbf{x})\right|_{\mathbf{x}=\mathbf{0}} + 	\left(\left.\partial_{x_0} \phi(\mathbf{x})\right|_{\mathbf{x}=\mathbf{0}}\right)^2 .\\
\eesplit
With the definitions
\besplit \label{EqC:h1h2Def}
	h_1^+(\mathbf{x}) &=\frac{1}{4}~\frac{\lambda^{\textnormal{fl}}_{-\mathbf{x}+\mathbf{k^0}}(-\beta)}{\lambda_{-\mathbf{x}+\mathbf{k^0}}-\lambda_{-\mathbf{x}}}, \\ 
	h_2^+(\mathbf{x}) &=\lambda^{\textnormal{fl}}_{-\mathbf{x}}(\beta)  \frac{e^{(\lambda_{-\mathbf{x}+\mathbf{k^0}}-\lambda_{-\mathbf{x}})\tau}-1}{\lambda_{-\mathbf{x}+\mathbf{k^0}}-\lambda_{-\mathbf{x}}}, \\ 
	h_1^-(\mathbf{x}) &=\frac{1}{4}~\frac{\lambda^{\textnormal{fl}}_{-\mathbf{x}-\mathbf{k^0}}(\beta)}{\lambda_{-\mathbf{x}-\mathbf{k^0}}-\lambda_{-\mathbf{x}}}, \\ 
	h_2^-(\mathbf{x}) &=\lambda^{\textnormal{fl}}_{-\mathbf{x}}(-\beta) \frac{e^{(\lambda_{-\mathbf{x}-\mathbf{k^0}}-\lambda_{-\mathbf{x}})\tau}-1}{\lambda_{-\mathbf{x}-\mathbf{k^0}}-\lambda_{-\mathbf{x}}},
\eesplit
function $h(\mathbf{x})$ can be written as
\besplit	\label{EqC:hStructure}
	h(\mathbf{x}) &= h_1^+(\mathbf{x})h_2^+(\mathbf{x}) + h_1^-(\mathbf{x})h_2^-(\mathbf{x}) \\ 
		      &-\left(h_1^+(\mathbf{x}) \lambda^{\textnormal{fl}}_{-\mathbf{x}}(\beta)+ h_1^-(\mathbf{x})\lambda^{\textnormal{fl}}_{-\mathbf{x}}(-\beta)\right)   \tau. 
\eesplit
Again, only the derivatives of terms $h_1^{\pm}(\mathbf{x})\lambda^{\textnormal{fl}}_{-\mathbf{x}}(\pm\beta)$ remain in the limit $\tau\to\infty$. The derivatives of the eigenvalues with respect to the first component of $\mathbf{x}$ are the same as those for the single oscillator
\besplit	\label{EqC:MomentsDerive}
	\partial_{x_0} \left.\lambda^{\textnormal{fl}}_{-\mathbf{x}+\mathbf{k^0}}(\beta)\right|_{\mathbf{x}=\mathbf{0}} &= \left. \frac{d}{dx_0} \lambda^{\textnormal{fl}}_{-x_0+k_0^0} \right|_{x_0=0}	\\
	\partial_{x_0} \left.\lambda^{\textnormal{df}}_{-\mathbf{x}+\mathbf{k^0}}\right|_{\mathbf{x}=\mathbf{0}} &= \left. \frac{d}{dx_0} \lambda^{\textnormal{df}}_{-x_0+k_0^0} \right|_{x_0=0}
\eesplit
given by Eq.~(\ref{EqB:MomentsDerive}) in Appendix B. Thus, functions $h_1^\pm(\mathbf{x})$ and their derivatives at zero are
\besplit
\label{EqC:h1ZeroLimits}
	& h_1^+(\mathbf{0}) =\frac{1}{4}~\frac{\lambda^{\textnormal{fl}}_{\mathbf{k^0}}(-\beta)}{\lambda_{\mathbf{k^0}}} 	\\ 
	&\partial_{x_0}\left.h_1^{+}(\mathbf{x})\right|_{\mathbf{x}=\mathbf{0}}  \\
	&=  i\frac{1}{4} \left[\frac{\lambda^{\textnormal{fl}}_{k^0_0}+L}{\lambda_{\mathbf{k^0}}} 
	 -\frac{\lambda^{\textnormal{fl}}_{\mathbf{k^0}}(-\beta)\left(\lambda^{\textnormal{fl}}_{k^0_0}+i 2\gamma \textnormal{Im}\left[\lambda^{\textnormal{fl}}_{k^0_0}\right]\right)}{\lambda^2_{\mathbf{k^0}}}\right].
	\raisetag{80pt}
\eesplit
as well as $h^-(\mathbf{0})=h^+(\mathbf{0})^*$ and $\partial_{x_0} h^-(\mathbf{0})= \partial_{x_0} h^+(\mathbf{0})^*$.
The mean frequency and phase diffusion constant of the first oscillator are then determined as
\besplit
\label{EqC:MeanOmegaRing}
	&\Omega = \frac{2\pi}{L} \lim_{\tau\to\infty} \frac{1}{\tau}\mathbb{E}[n_0] = 2\pi - \varepsilon^2 \frac{2\pi}{L}\lim_{\tau\to\infty} \frac{1}{\tau}i\partial_{x_0}\left.h'(\mathbf{x})\right|_{\mathbf{x}=\mathbf{0}}	 
	\\ 
	&= 2\pi + \varepsilon^2 \frac{2\pi}{L} i\left. \left[h^+_1(\mathbf{x})\partial_{x_0}\lambda^{\textnormal{fl}}_{-\mathbf{x}}(\beta)+h^-_1(\mathbf{x})\partial_{x_0}\lambda^{\textnormal{fl}}_{-\mathbf{x}}(-\beta) \right] \right|_{\mathbf{x}=0}  \\  
	&= 2\pi \left( 1 - \varepsilon^2 \frac{1}{4}\left[\frac{\lambda^{\textnormal{fl}}_{\mathbf{k^0}}(-\beta)}{\lambda_{\mathbf{k^0}}} 
	                                                + \frac{\lambda^{\textnormal{fl}}_{-\mathbf{k^0}}(\beta)}{\lambda_{-\mathbf{k^0}}}\right]\right),
	\raisetag{20pt}
\eesplit
%
%
\besplit	\label{EqC:MeanD0Ring}
      D & = \left(\frac{2\pi}{L}\right)^2 \lim_{\tau\to\infty} \frac{1}{2\tau} \textnormal{Var}[n] \\
	& = \frac{2\pi^2}{L} (1+2\gamma) - \varepsilon^2 \left(\frac{2\pi}{L}\right)^2\lim_{\tau\to\infty} \frac{1}{2\tau}\partial^2_0 \left.h(\mathbf{x})\right|_{\mathbf{x}=\mathbf{0}}
	\raisetag{50pt},
\eesplit
so that
\\ \\
\bmul \label{EqC:MeanPhaseDiffRing}
	\frac{D}{D_0} =	1 + \varepsilon^2 \frac{1}{L(1+2\gamma)} 
	\left[ 
	    \vphantom{- \frac{2 (\lambda^{\textnormal{fl}}_{k^0_0}+L)+\lambda^{\textnormal{fl}}_{\mathbf{k^0}}(-\beta)}{\lambda_{\mathbf{k^0}}}}
		      2 \partial_{x_0} h^+_1(\mathbf{x})\partial_{x_0}\lambda_{-\mathbf{x}}(\beta) \right. \\
		      \left.\left. 
	    \vphantom{- \frac{2 (\lambda^{\textnormal{fl}}_{k^0_0}+L)+\lambda^{\textnormal{fl}}_{\mathbf{k^0}}(-\beta)}{\lambda_{\mathbf{k^0}}}}
		      + h^+_1(\mathbf{x})\partial^2_0\lambda_{-\mathbf{x}}(\beta) +  c.c.\right]\right|_{\mathbf{x}=\mathbf{0}}	 \\
	\shoveleft = 1 +\varepsilon^2  \frac{1}{(1+2\gamma)} \left.\left[
	    \vphantom{- \frac{2 (\lambda^{\textnormal{fl}}_{k^0_0}+L)+\lambda^{\textnormal{fl}}_{\mathbf{k^0}}(-\beta)}{\lambda_{\mathbf{k^0}}}}
	2i\partial_{x_0} h^+_1(\mathbf{x}) - h^+_1(\mathbf{x}) +  c.c.\right]\right|_{\mathbf{x}=\mathbf{0}}	\\
	\shoveleft = 1+\varepsilon^2 \frac{1}{4(1+2\gamma)} \left[2 \frac{\lambda^{\textnormal{fl}}_{\mathbf{k^0}}(-\beta)\left(\lambda^{\textnormal{fl}}_{k^0_0}+i 2\gamma \textnormal{Im}\left[\lambda^{\textnormal{fl}}_{k^0_0}\right]\right)}{\lambda^2_{\mathbf{k^0}}} \right. \\
	\left. - \frac{2 (\lambda^{\textnormal{fl}}_{k^0_0}+L)+\lambda^{\textnormal{fl}}_{\mathbf{k^0}}(-\beta)}{\lambda_{\mathbf{k^0}}}  + c.c.
	\right]
\emul
and for the P\'eclet number
\bmul
\label{EqC:MeanCoherenceRing}
	\frac{c}{c_0} = 1 + \varepsilon^2 \left(\frac{\Omega}{\omega_0}-\frac{D}{D_0}\right) \\
	\shoveleft = 1 + \varepsilon^2 \frac{1}{2(1+2\gamma)}\left[ \frac{\lambda^{\textnormal{fl}}_{k_0^0} - \gamma \lambda^{\textnormal{fl}}_\mathbf{k^0}(-\beta) + L}{\lambda_\mathbf{k^0}}\right. \\
	\left.
		- \frac{\lambda^{\textnormal{fl}}_{\mathbf{k^0}}(-\beta)\left(\lambda^{\textnormal{fl}}_{k^0_0}+i 2\gamma \textnormal{Im}\left[\lambda^{\textnormal{fl}}_{k^0_0}\right]\right)}{\lambda^2_{\mathbf{k^0}}} + c.c.
	\right].
\emul
\\[10ex]
\end{document}